%% file: ANA-TOPQ-2021-11-PAPER.tex
\pdfoutput=1
\documentclass[PAPER, atlasdraft=false, UKenglish, cernpreprint, texmf, orcidlogo]{atlasdoc}
\usepackage[subcaption]{atlaspackage}
\usepackage{atlasbiblatex}

\usepackage[jetetmiss=true]{atlasphysics}
\graphicspath{{logos/}{figures/}}

\usepackage{ANA-TOPQ-2021-11-PAPER-defs}
\usepackage{tikz-feynman}
\usepackage{multirow}
\usepackage{makecell}

\AtlasTitle{Measurement of the top-quark Yukawa coupling from $t \overline{t}$ production in the lepton+jets final state using $pp$ collisions at $\sqrt{s} = 13$~TeV with the ATLAS detector}

\AtlasAbstract{%
The top-quark Yukawa coupling is extracted from the distribution of the
top-quark pair ($t \overline{t}$) invariant mass in proton--proton collisions using
$140\,\mathrm{fb}^{-1}$ of data at $\sqrt{s} = 13$~TeV collected in
2015--2018 by the ATLAS experiment at the Large Hadron Collider.  In the region near the production
threshold, the $t \overline{t}$ invariant mass spectrum is sensitive to electroweak
virtual corrections, including contributions from Higgs boson exchange,
thereby providing sensitivity to the top-quark Yukawa coupling. This is the first
measurement in ATLAS that aims to obtain this coupling exploiting this approach.
The $t\overline{t}$ system is reconstructed in the single-lepton final state,
requiring exactly one isolated electron or muon and at least four jets
with at least two identified as originating from $b$-quarks.  The
measured Yukawa coupling is found to be in good agreement with the Standard
Model prediction.  An upper limit on the top-quark Yukawa coupling
strength of $Y_t < 2.1$ relative to the Standard Model prediction is
observed at 95\% confidence level, consistent with the expected
sensitivity.  }

\PreprintIdNumber{CERN-EP-2025-216}

\AtlasJournalRef{\JHEP 01 (2026) 117}
\AtlasDOI{10.1007/JHEP01(2026)117}

\hypersetup{pdftitle={ATLAS document},pdfauthor={The ATLAS Collaboration}}

\usepackage{graphicx}
\begin{document}

\maketitle

\tableofcontents

\section{Introduction}
\label{sec:intro}

% The next lines are included from the .//analysis_main/Introduction.tex input file
In the Standard Model (SM), fermions obtain their masses ($m_f$) via
spontaneous symmetry breaking %
as described by the Englert-Brout-Higgs mechanism~\cite{Englert:1964et,Higgs:1964ia,Higgs:1964pj,Guralnik:1964eu,Higgs:1966ev,Kibble:1967sv}.
The coupling between the Higgs-boson and fermion fields is described by a Yukawa
coupling $g_f = \sqrt{2} m_f / v$ where $v=246.22$~\GeV~\cite{ParticleDataGroup:2024cfk} is the vacuum
expectation value of the Higgs field.

Since the top-quark mass is heavier than half the mass of the Higgs boson, the top
Yukawa coupling ($g_t$) is the only fermion Yukawa coupling that cannot be
measured directly from Higgs boson decays.
The most model-independent direct measurement uses the $\ttbar{}H$
process~\cite{HIGG-2020-24,HIGG-2021-23,CMS-HIG-22-001} where the result
depends on the total width of the Higgs boson. Complementary
information can be obtained from other processes, such as
$gg \rightarrow H$ or $H \rightarrow \gamma \gamma$~\cite{HIGG-2021-23,CMS-HIG-22-001} where the top
quarks appear in a loop, or from four-top-quark
production~\cite{TOPQ-2021-08,CMS-TOP-18-003}, where off-shell Higgs bosons
can be produced. In addition, virtual corrections involving Higgs bosons also
have a significant impact on the \ttbar production
cross-section~\cite{Kuhn:2013zoa}.  The numerically relevant corrections
include a Higgs boson exchange between any two top-quark lines.
In case the Standard Model holds, all measurements result in a coupling
strength \yt ($= g_t/g_t^{\mathrm{SM}}$)\footnote{
$\yt$ is identical to the coupling strength modifier $\kappa_t$ as defined
e.g.~in Ref.~\cite{HIGG-2021-23,CMS-HIG-22-001}}
compatible with unity. However, if physics beyond the SM exists, it could affect
each of these measurements differently, making them complementary.

\begin{figure}[htbp]
\centering
\vspace{4mm}
\begin{subfigure}[l]{.43\textwidth}
\includegraphics[width=1\textwidth]{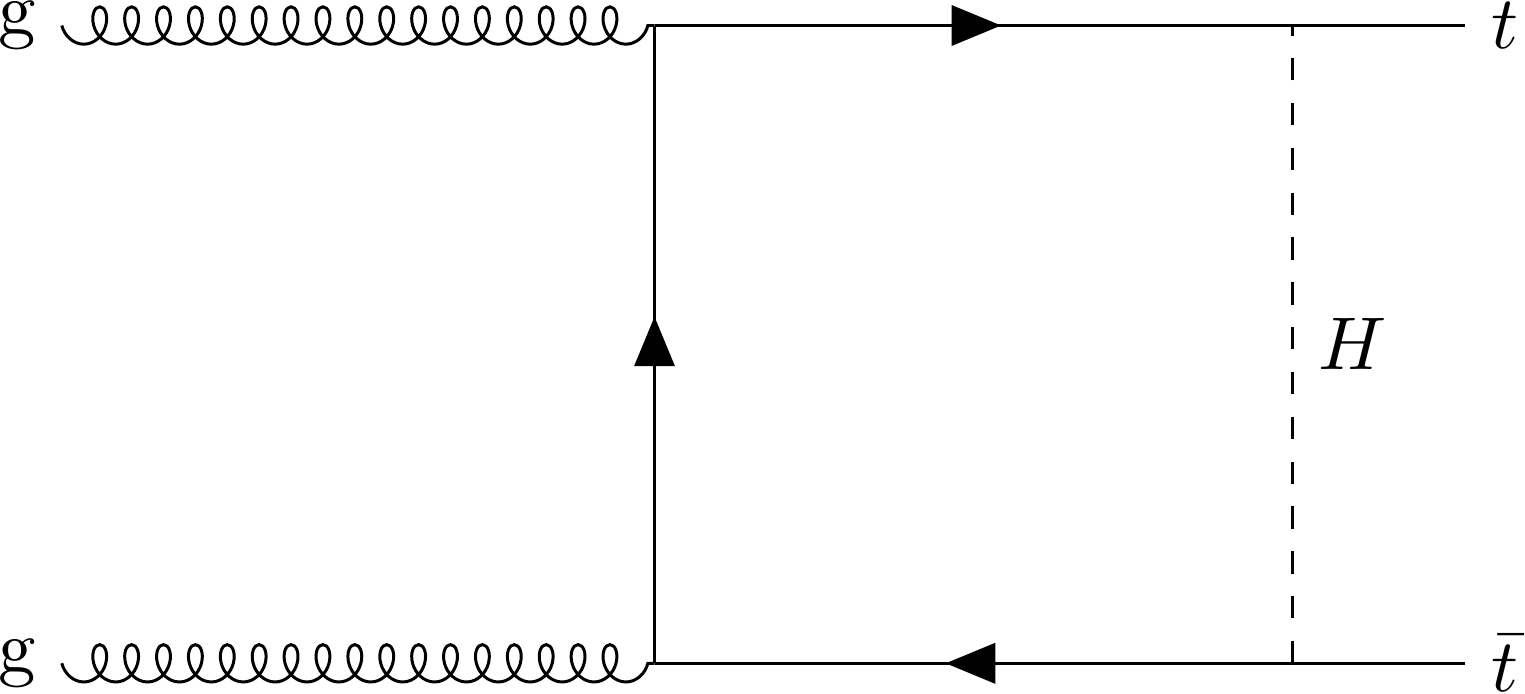}
\label{fig:gg}
\caption{}
\end{subfigure}
\hspace{1cm}
\begin{subfigure}[r]{.43\textwidth}
\includegraphics[width=1\textwidth]{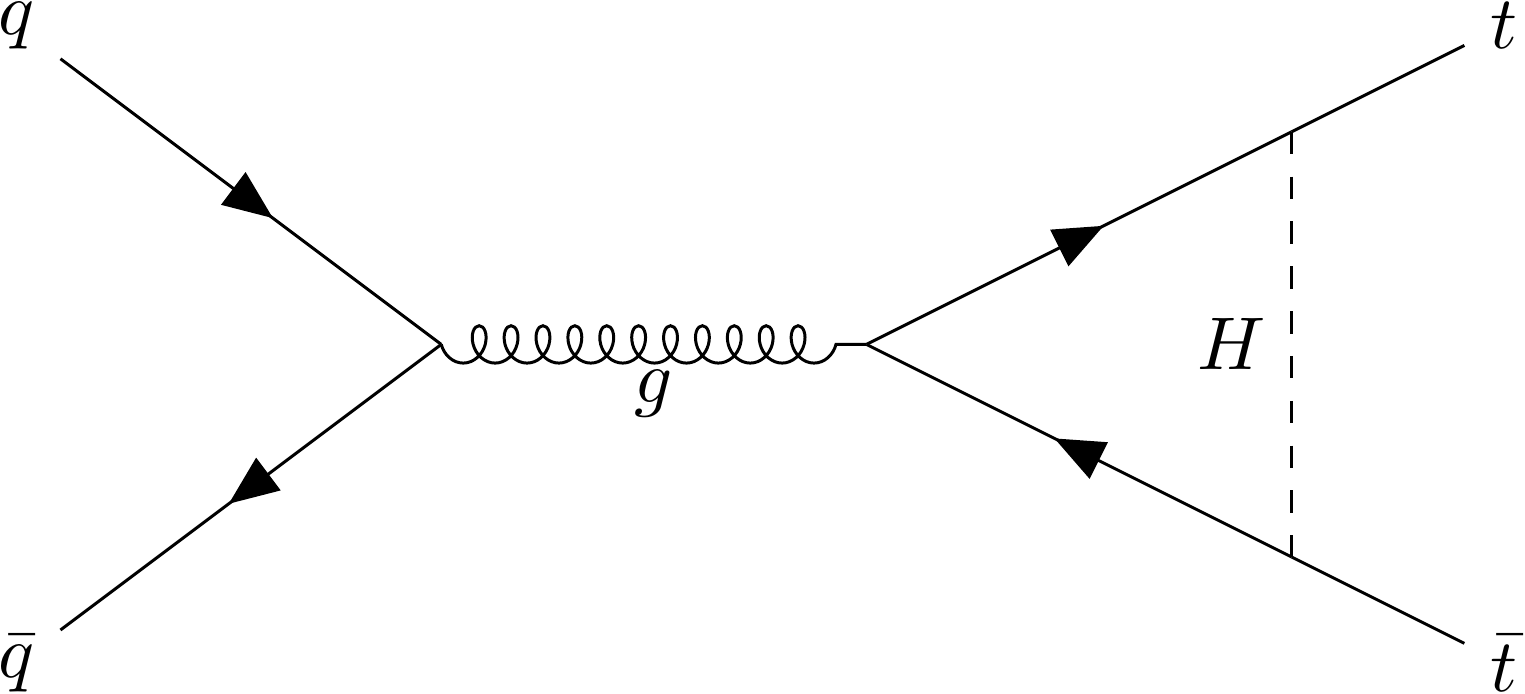}
\label{fig:qq}
\caption{}
\end{subfigure}
\caption{Example Feynman diagrams for (a) \gg and (b) \qqbar
\ttbar production with virtual Higgs boson exchange.}
\label{fig:feynman_diagrams}
\end{figure}

Examples of Feynman diagrams for virtual Higgs-boson corrections to \ttbar
production are shown in Figure~\ref{fig:feynman_diagrams}.
The corresponding amplitudes depend on
\ytsq and are sizeable mainly near the
\ttbar production threshold, where the top and anti-top quarks have a small
relative velocity. The largest sensitivity is therefore expected in this kinematic region.
Interfering with the Born level amplitudes, the corrections to
the cross-section are also quadratic in \yt.
One can therefore extract \ytsq from a
measurement of the \ttbar production cross-section as a function of the
\ttbar invariant mass, \mtt.
The corrections also depend on the top quark
scattering angle in the \ttbar rest frame, \costhst,
although this dependence is weak in the kinematic region most relevant for
the measurement.
The method is insensitive to the sign of the Yukawa coupling.
Therefore $\yt$ is always assumed to be $|\yt|$ in this paper.
While the sensitivity to \yt from virtual corrections is not expected to
match that of direct measurements, such as those from $\ttbar H$
production~\cite{HIGG-2020-24,HIGG-2021-23,CMS-HIG-22-001}, the approach remains highly valuable. As demonstrated in
Ref.~\cite{Maltoni:2024wyh}, contributions from new physics, such as a
top-philic scalar, can modify the virtual corrections in a way that mimics a deviation in \yt, potentially affecting indirect
and direct extractions differently.
Similar models are tested by searches in four-top final
states~\cite{EXOT-2022-13} and by searches for new Higgs bosons decaying into
\ttbar~\cite{EXOT-2020-25}.

The CMS Collaboration already performed this analysis in the single lepton
channel~\cite{CMS-TOP-17-004} and in the
dilepton channel~\cite{CMS-TOP-19-008}, %
obtaining observed 95\% confidence level upper limits on \yt{} of 1.67 and 1.54, respectively.

In this paper, the first ATLAS measurement of \yt extracted from top-quark
pair production is presented. The measurement uses the ATLAS data at
$\sqrt{s} = 13~\TeV$ with an integrated luminosity of $140$~\ifb.  The
single-lepton channel is analysed where the \ttbar invariant mass can be
fully reconstructed and the background is low. Electroweak (EW) corrections
for variable \yt are calculated in~\cite{Kuhn:2006vh,Kuhn:2005it,Kuhn:2013zoa} and implemented in
\hathor~\cite{Hathor-21-b3}, a program that calculates the \ttbar cross-section
at parton level in leading-order QCD. They are used to reweight
fully-simulated Monte Carlo (MC) events. The weighting is done as a
function of the initial-state parton flavour, the generated \mtt and \costhst.
This allows to predict the \mtt spectra as a function of \yt.
The measured \mtt spectrum is then fitted in a profile likelihood fit, using
these predictions with \ytsq as the parameter of interest, together with a
set of nuisance parameters representing the systematic uncertainties.

A brief description of the ATLAS detector is provided in Section~\ref{sec:detector},
followed by an overview of the data and MC samples in Section~\ref{sec:dataMCsamples},
with Section~\ref{sec:electroweakCorr} describing the electroweak corrections applied in the analysis.
Section~\ref{sec:eventReco} explains the event reconstruction and selection procedures, including the estimate of the fake-lepton background. The systematic uncertainties are discussed in Section~\ref{sec:syst}, and the fit strategy is outlined in Section~\ref{sec:fit}. Finally, the results are presented in Section~\ref{sec:result}, with conclusions summarised in Section~\ref{sec:conclusion}.

% End of text imported from the .//analysis_main/Introduction.tex input file

%
\section{ATLAS detector}
\label{sec:detector}

% The next lines are included from the .//analysis_main/Atlas.tex input file

%

%

%
\newcommand{\AtlasCoordFootnote}{%
ATLAS uses a right-handed coordinate system with its origin at the nominal interaction point (IP)
in the centre of the detector and the \(z\)-axis along the beam pipe.
The \(x\)-axis points from the IP to the centre of the LHC ring,
and the \(y\)-axis points upwards.
Polar coordinates \((r,\phi)\) are used in the transverse plane,
\(\phi\) being the azimuthal angle around the \(z\)-axis.
The pseudorapidity is defined in terms of the polar angle \(\theta\) as \(\eta = -\ln \tan(\theta/2)\) and is equal to the rapidity
$ y = \frac{1}{2} \ln \left( \frac{E + p_z}{E - p_z} \right) $ in the relativistic limit.
Angular distance is measured in units of \(\DeltaR \equiv \sqrt{(\Delta y)^{2} + (\Delta\phi)^{2}}\).}

The ATLAS detector~\cite{PERF-2007-01} at the Large Hadron Collider (LHC)~\cite{Evans:2008zzb} covers nearly the entire solid angle around the collision point.\footnote{\AtlasCoordFootnote}
It consists of an inner tracking detector surrounded by a thin superconducting solenoid, electromagnetic and hadronic calorimeters,
and a muon spectrometer incorporating three large superconducting air-core toroidal magnets.

The inner-detector system (ID) is immersed in a \qty{2}{\tesla} axial magnetic field
and provides charged-particle tracking in the range \(|\eta| < 2.5\).
The high-granularity silicon pixel detector covers the vertex region and typically provides four measurements per track,
the first hit generally being in the insertable B-layer (IBL) installed before Run~2~\cite{ATLAS-TDR-19,PIX-2018-001}.
It is followed by the SemiConductor Tracker (SCT), which usually provides eight measurements per track.
These silicon detectors are complemented by the transition radiation tracker (TRT),
which enables radially extended track reconstruction up to \(|\eta| = 2.0\).
The TRT also provides electron identification information
based on the fraction of hits (typically 30 in total) above a higher energy-deposit threshold corresponding to transition radiation.

The calorimeter system covers the pseudorapidity range \(|\eta| < 4.9\).
Within the region \(|\eta|< 3.2\), electromagnetic calorimetry is provided by barrel and
endcap high-granularity lead/liquid-argon (LAr) calorimeters,
with an additional thin LAr presampler covering \(|\eta| < 1.8\)
to correct for energy loss in material upstream of the calorimeters.
Hadronic calorimetry is provided by the steel/scintillator-tile calorimeter,
segmented into three barrel structures within \(|\eta| < 1.7\), and two copper/LAr hadronic endcap calorimeters.
The solid angle coverage is completed with forward copper/LAr and tungsten/LAr calorimeter modules
optimised for electromagnetic and hadronic energy measurements respectively.

The muon spectrometer (MS) comprises separate trigger and
high-precision tracking chambers measuring the deflection of muons in a magnetic field generated by the superconducting air-core toroidal magnets.
The field integral of the toroids ranges between \num{2.0} and \qty{6.0}{\tesla\metre}
across most of the detector.
Three layers of precision chambers, each consisting of layers of monitored drift tubes, cover the region \(|\eta| < 2.7\),
complemented by cathode-strip chambers in the forward region, where the background is highest.
The muon trigger system covers the range \(|\eta| < 2.4\) with resistive-plate chambers in the barrel, and thin-gap chambers in the endcap regions.

The luminosity is measured mainly by the LUCID--2~\cite{LUCID2} detector that records Cherenkov light produced in the quartz windows of photomultipliers located close to the beampipe.

Events are selected by the first-level trigger system implemented in custom hardware,
followed by selections made by algorithms implemented in software in the high-level trigger~\cite{TRIG-2016-01}.
The first-level trigger accepts events from the \qty{40}{\MHz} bunch crossings at a rate below \qty{100}{\kHz},
which the high-level trigger further reduces in order to record complete events to disk at about \qty{1.25}{\kHz}.

A software suite~\cite{SOFT-2022-02} is used in data simulation, in the reconstruction
and analysis of real and simulated data, in detector operations, and in the trigger and data acquisition
systems of the experiment.

% End of text imported from the .//analysis_main/Atlas.tex input file

%
\section{Data and simulation samples}
\label{sec:dataMCsamples}

% The next lines are included from the .//analysis_main/DataMCSamples.tex input file
Proton--proton (\pp) collisions at $\rts = 13$~\TeV
collected in 2015--2018 by the ATLAS experiment are analysed~\cite{DAPR-2018-01}.
Only the events where all components of the ATLAS detector were
fully functional are selected, resulting in a total dataset corresponding to an integrated luminosity of $140$~\ifb.
Selected events are required to be triggered by one of the unprescaled
single-lepton triggers which are further described in Section~\ref{subsec:evtsel}.

MC simulated event samples are used to estimate the signal and background contributions
containing prompt leptons.
Backgrounds containing misidentified or non-prompt leptons
stemming e.g.~from heavy-flavour decays are determined
using data-driven approach as detailed in Section~\ref{sub:fakes}.

The ATLAS simulation infrastructure~\cite{SOFT-2010-01} is used for all the simulation samples, with the detector response simulated using the GEANT~\cite{Agostinelli:2002hh} framework.
For the estimate of several signal modelling uncertainties, a fast simulation,
which utilises parametrisations of the hadronic showers in the EM and hadronic calorimeters
to speed up the simulation~\cite{ATL-PHYS-PUB-2010-013}, is used.
All simulation samples are processed with the same reconstruction software as the data samples.

To account for additional \pp interactions from the same or close-by bunch
crossings (pile-up), a set of minimum-bias
interactions generated with \PYTHIA[8.186]~\cite{Sjostrand:2007gs} using the \NNPDF[2.3lo] parton distribution function~(PDF) set~\cite{Ball:2012cx} with the A3 set of tuned parameters (tune)~\cite{ATL-PHYS-PUB-2016-017}
is superimposed to the hard scattering events.
Furthermore, the simulated events are reweighted according to the number of additional \pp interactions per bunch crossing to match the pile-up conditions of each dataset corresponding to the 2015--2018 years of data-taking.
Corrections are applied to simulated events to improve agreement between the data and the simulation samples in the object identifications, efficiencies, energy scales, and energy resolutions, as described in Section~\ref{subsec:objdefs}.

Heavy-flavour decays are modelled with the \textsc{EvtGen} program~\cite{Lange:2001uf} in all parton shower generators, with the exception of the \Sherpa~\cite{Gleisberg:2008ta} generator.

\subsection{Signal modelling}
\label{subsec:signal}

The nominal signal \ttbar MC sample is simulated using \POWHEGBOX-v2~\cite{Nason:2004rx,Frixione:2007vw,Alioli:2010xd}
which is based on next-to-leading-order (NLO) QCD matrix element (ME) calculations.
The calculation for the ME uses the \NNPDF[3.0nlo] PDF set~\cite{Ball:2014uwa}
with a top-quark mass ($m_t$) of 172.5~\gev.
The \hdamp parameter that controls the emission of the
first gluon is set to $1.5m_t$.
The renormalisation (\muR) and factorisation (\muF) scales are defined by the functional form $\sqrt{m_t^2 + \pTtop^2}$, where \pTtop is the transverse momentum of the top quark.
The ME generator is interfaced with \PYTHIA[8.230]~\cite{Sjostrand:2014zea} which simulates
parton shower (PS), fragmentation, hadronisation, and the underlying event.
The A14 tune~\cite{ATL-PHYS-PUB-2014-021} together with
\NNPDF[2.3lo] PDF set is applied for \PYTHIA[8] showering.
The \pthard parameter which impacts the matching of the ME to PS, is set to zero~\cite{	ATL-PHYS-PUB-2023-029}.

A number of signal modelling uncertainties are estimated using alternative MC samples.
To compare the effect of different PS and hadronisation modelling,
a sample produced with \POWHEGBOX-v2 interfaced with \HERWIG[7.2.1]~\cite{Bellm:2015jjp,Bahr:2008pv} with the \MMHT[lo] PDF set~\cite{Harland-Lang:2014zoa} and the default set of tuned parameters, is used. The \POWHEG settings are the same as in the nominal sample.
For the estimate of the \hdamp uncertainty, an alternative sample with the same generator and settings as the nominal \ttbar{} sample and the \hdamp parameter doubled to $3m_t$, is used.
Similarly, alternative samples based on the nominal sample with $m_t$ varied to 172~\gev and 173~\gev, respectively, are used to estimate the impact of the assumption on $m_t$ on the analysis.
Finally, to estimate the uncertainty in the ME-to-PS matching, a MC sample produced with the same generator and the same settings as the default \ttbar sample is used, except for the \pthard parameter setting, which is set to $\pthard=1$~\cite{ATL-PHYS-PUB-2023-029}.
The \ttbar production threshold is potentially sensitive to the modelling of off-shell effects and top-quark decay.
To estimate the uncertainty in the modelling of these effects, an alternative sample based on the nominal \ttbar sample, but with the \MADSPIN generator interfaced to \POWHEGBOX-v2 to simulate the top-quark decay~\cite{Frixione:2007zp,Artoisenet:2012st}, is used.

All \ttbar{} MC samples are normalised
to next-to-next-to-leading-order (NNLO) cross-section including the resummation of soft gluon emissions at next-to-next-to-leading-logarithmic
(NNLL) accuracy using \textsc{Top++2.0}~\cite{Czakon:2011xx}.
The resulting cross-section for the \ttbar process is $\sigma_{\ttbar} = 834^{+21}_{-30}(\text{scale})\pm21(\text{PDF}+\alpha_S)$~pb for $m_t=172.5~\GeV$~\cite{Beneke:2011mq,Cacciari:2011hy,Baernreuther:2012ws,Czakon:2012zr,Czakon:2012pz,Czakon:2013goa}.

Additionally, a dedicated fixed-order NNLO QCD prediction~\cite{Catani:2019iny,Catani:2019hip} is used to define an uncertainty due to missing higher-order corrections.
It is calculated using the \textsc{MATRIX} tool~\cite{Grazzini:2017mhc,Denner:2016kdg,Cascioli:2011va,Buccioni:2019sur,Buccioni:2017yxi,Barnreuther:2013qvf}, using the \NNPDF[3.0nnlo] PDF set~\cite{Ball:2014uwa} and assuming $m_t = 172.5~\GeV$.
The functional form of the \muR and \muF scales is set to $\HT/4$, where $\HT = \sqrt{m_t^2 + \pTtop^2} + \sqrt{m_t^2 + \pTtbar^2}$, based on recommendations in Ref.~\cite{Mazzitelli_2022}.
The prediction is calculated in terms of a two-dimensional distribution of \mtt and the cosine of the angle between the top-quark momentum boosted to the \ttbar{} rest-frame and the momentum of the \ttbar{} system in the laboratory frame.
Subsequently, the nominal \ttbar MC sample is reweighted using this prediction to create a systematically varied alternative sample.

\subsection{Background modelling}
\label{subsec:backgrounds}

In the \ttbar threshold region, the formation of a colour-singlet \ttbar{} quasi-bound state, commonly referred to as the \emph{toponium}, is expected, consistent with the recent CMS~\cite{CMS-TOP-24-007} observation indicating an excess in the \ttbar threshold region.
The toponium contribution is described using a simplified model~\cite{Fuks:2021xje,maltoni2024quantumdetectionnewphysics}, as a pseudo-scalar $s$-channel resonance, since the dominant contribution to the $gg\rightarrow \ttbar{}$ colour-singlet quasi-bound state comes from pseudo-scalar $J=0$ states~\cite{Hagiwara:2008df}.
The width of the resonance is set to 2.8~\GeV and the mass of the resonance to 343~\GeV, assuming that the toponium quasi-bound state has twice the mass of the top quark minus the binding energy of approximately 2~\GeV~\cite{Hagiwara:2008df,threshold_Sumino_2010}. The toponium events are simulated using \textsc{MadGraph}~3.5.5 with the \NNPDF[3.0nlo] PDF set, interfaced to \PYTHIA[8.3]~\cite{Bierlich:2022pfr} for PS and hadronisation modelling.
At present, the impact of electroweak corrections on the toponium is unknown and thus neglected.
Therefore, the toponium contribution is considered as a background process. It is normalised to a non-relativistic perturbative QCD prediction of \SI{6.43}{\pico\barn}~\cite{Fuks:2021xje,threshold_Sumino_2010}.

Single-top quark processes are split into $s$-channel, $t$-channel and $tW$-channel contributions.
They are simulated using the same setup as the nominal \ttbar sample, using the five-flavour scheme in the PDF set, with the exception of the $t$-channel process, which is generated using the four-flavour scheme.
The overlap between \ttbar and single-top $Wt$ final states is removed using the diagram removal (DR) technique~\cite{Frixione:2008yi} and an additional sample using diagram subtraction (DS)~\cite{ATL-PHYS-PUB-2016-020,Frixione:2008yi} is used for the estimate of the corresponding modelling uncertainty.
The single-top samples are normalised to NNLO QCD predictions for the $s$- and $t$-channels~\cite{Campbell:2020fhf,Kidonakis:2010tc} and to approximate N$^3$LO QCD prediction for the $tW$ channel~\cite{Kidonakis:2021vob}

Several background processes are simulated using different versions of the \SHERPA~\cite{Bothmann:2019yzt} generator, with the details outlined below. \SHERPA includes both the ME, PS and hadronisation modelling. The matrix elements are computed using \Comix{}~\cite{Gleisberg:2008fv} and \OpenLoops{}~\cite{Cascioli:2011va}. The \NNPDF[3.0nlo] PDF set is used with a dedicated tune provided by the \SHERPA authors. The ME is merged with the \SHERPA PS using the \MEPS{} prescription~\cite{Hoeche:2012yf}.

Events with a $Z$ or $W$ boson in association with additional jets are
simulated with the \SHERPA[2.2.11] generator.
The matrix elements for up to two additional partons are calculated at NLO QCD precision, and at LO QCD for up to four additional partons.
The samples are normalised to the NNLO QCD prediction~\cite{ATL-PHYS-PUB-2023-026,Anastasiou:2003ds}.

Diboson ($WW$/$WZ$/$ZZ$) samples are simulated using \SHERPA[2.2.1]. The matrix elements for up to one additional parton are calculated at NLO QCD precision, and at LO QCD accuracy for up to three additional partons.
The samples are normalised to the NLO QCD theoretical cross-sections~\cite{Campbell:1999ah}.

A minor contribution to the total background originates from boson-associated \ttbar{} production.
The production of $\ttbar{}Z$ events is modelled using the
\MGNLO[2.3.3]~\cite{Alwall:2014hca} generator at NLO with the
\NNPDF[3.0nlo] PDF set.
The events are interfaced to \PYTHIA[8.210]
using the A14 tune and the
\NNPDF[2.3lo] PDF set.
The $\ttbar{}W$ process is simulated using \SHERPA[2.2.10]~\cite{Bothmann:2019yzt}. The matrix element is calculated at NLO QCD for up to one additional parton and up to two additional partons at LO QCD.
The production of $\ttbar{}H$ events is modelled using the
\POWHEGBOX[v2]~\cite{Hartanto:2015uka}
generator at NLO with the \NNPDF[3.0nlo] PDF set, interfaced to \PYTHIA[8.230] using the A14 tune and the \NNPDF[2.3lo] PDF set.
The samples corresponding to all three processes are normalised to the NLO QCD+EW predictions~\cite{deFlorian:2016spz}.

% End of text imported from the .//analysis_main/DataMCSamples.tex input file

\subsection{Electroweak corrections}
\label{sec:electroweakCorr}

% The next lines are included from the .//analysis_main/ElectroweakCorr.tex input file
To obtain the signal samples for different \yt values, a reweighting technique is used on the nominal \ttbar sample, which is based on a leading-order calculation for the full electroweak corrections where \yt is a free parameter.

The electroweak corrections with variable \yt as calculated in
\cite{Kuhn:2006vh,Kuhn:2005it,Kuhn:2013zoa} and implemented in \hathor
2.1-b3~\cite{Hathor-21-b3} are used.  The program evaluates the leading-order electroweak corrections to the Born level
$q\overline{q} \rightarrow \ttbar$ and $gg \rightarrow \ttbar$
cross-sections as a function of \mtt{} and \costhst{} assuming stable top
quarks.  More recent predictions for the electroweak corrections exist,
including additional amplitudes with the $s$-channel Higgs boson exchange
diagrams~\cite{Maltoni:2024wyh}, which are not considered in the predictions
obtained from \hathor.  These amplitudes have a sizeable impact on CP-odd
extensions of the Yukawa interaction. However, for the measurement presented
here, which assumes a CP-even \yt, their impact is
negligible~\cite{Maltoni:2024wyh}. The impact of a modified \yt on the
\ttbar production cross-section is evaluated in two steps. \hathor
predictions as functions of \mtt and \costhst are first parameterised with
smooth functions. These are then used to compute event weights, defined as
the ratio of predictions for a given \yt to those without EW corrections
applied. These weights are applied to the events generated with the nominal
\POWPY setup.  The weights depend on the parton-level \mtt and
\costhst, and initial-state parton flavour.  This corresponds to the
multiplicative approach combining electroweak and QCD corrections.  At
$\sqrt{s} = 13~\TeV$, about 2\% of the \POWHEG events originate from a
\qg{} or \qbarg{} initial state. For them the EW corrections corresponding
to the \gg{} initial state are used for reweighting.
This is a valid approach as in the region near the threshold, where the
sensitivity is the largest, the correction depends very little on the
initial state.  The top-quark mass is set to $m_t = 172.5~\GeV$,
consistent with the signal Monte Carlo samples used in the analysis.  The
weights are calculated as a function of \yt{}. For \gg{} initial states, in
all Feynman diagrams containing virtual Higgs boson corrections, the Higgs
boson is connected to two top-quark lines. For \qqbar{} initial states, the
same is true for the numerically dominant diagrams. For this reason, the
corrections are proportional to \ytsq{}.  The \yt{}-dependent electroweak
correction as a function of \mtt{} are shown in Figure~\ref{fig:ewcorr} for
the quark-induced as well as for the gluon-induced processes.

\begin{figure}[htbp]
\centering
\hspace*{-3mm}
\begin{subfigure}[l]{.5\textwidth}
\includegraphics[width=1\textwidth]{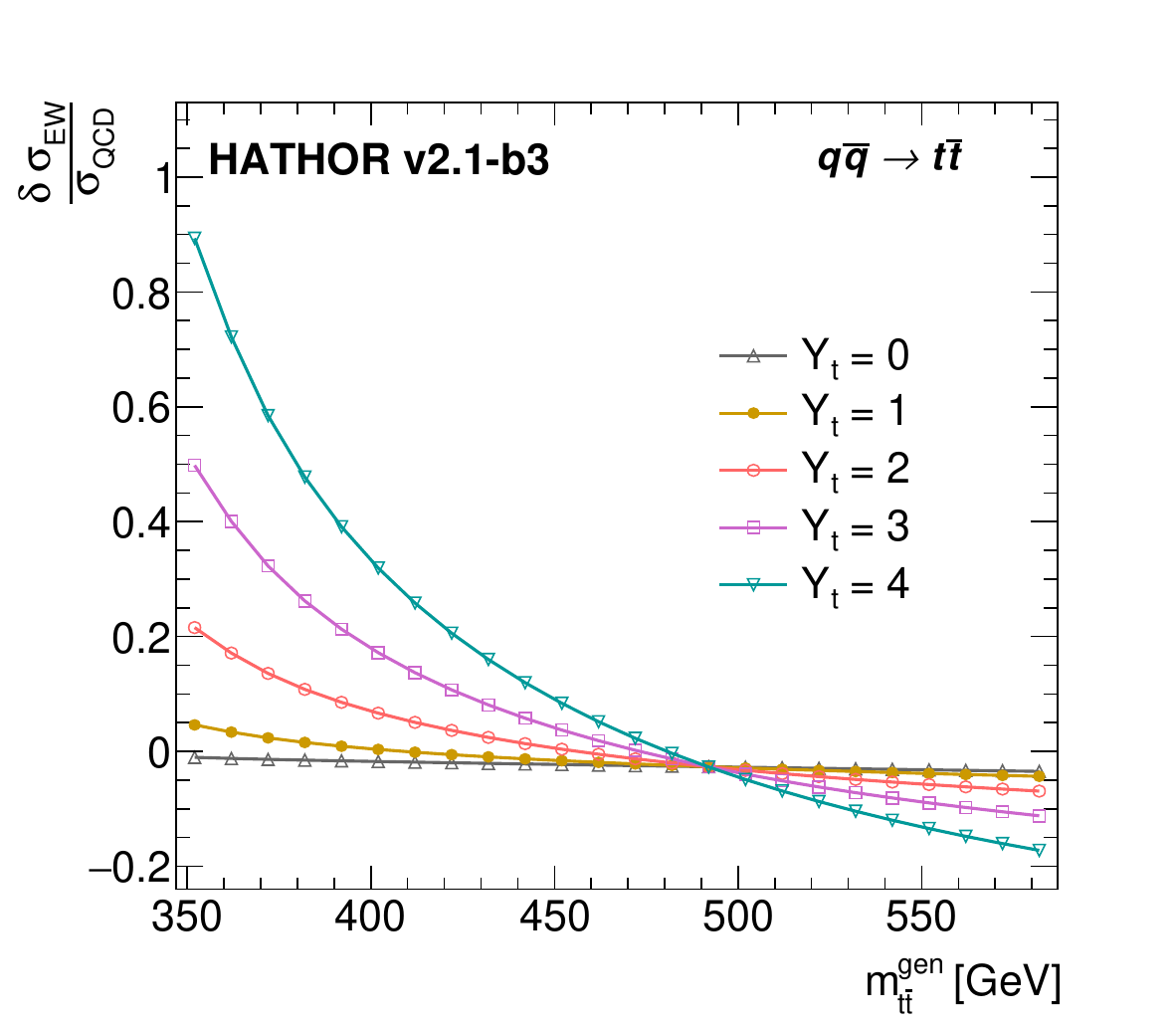}
\caption{}
\end{subfigure}
\centering
\hspace*{-3mm}
\begin{subfigure}[l]{.5\textwidth}
\includegraphics[width=1\textwidth]{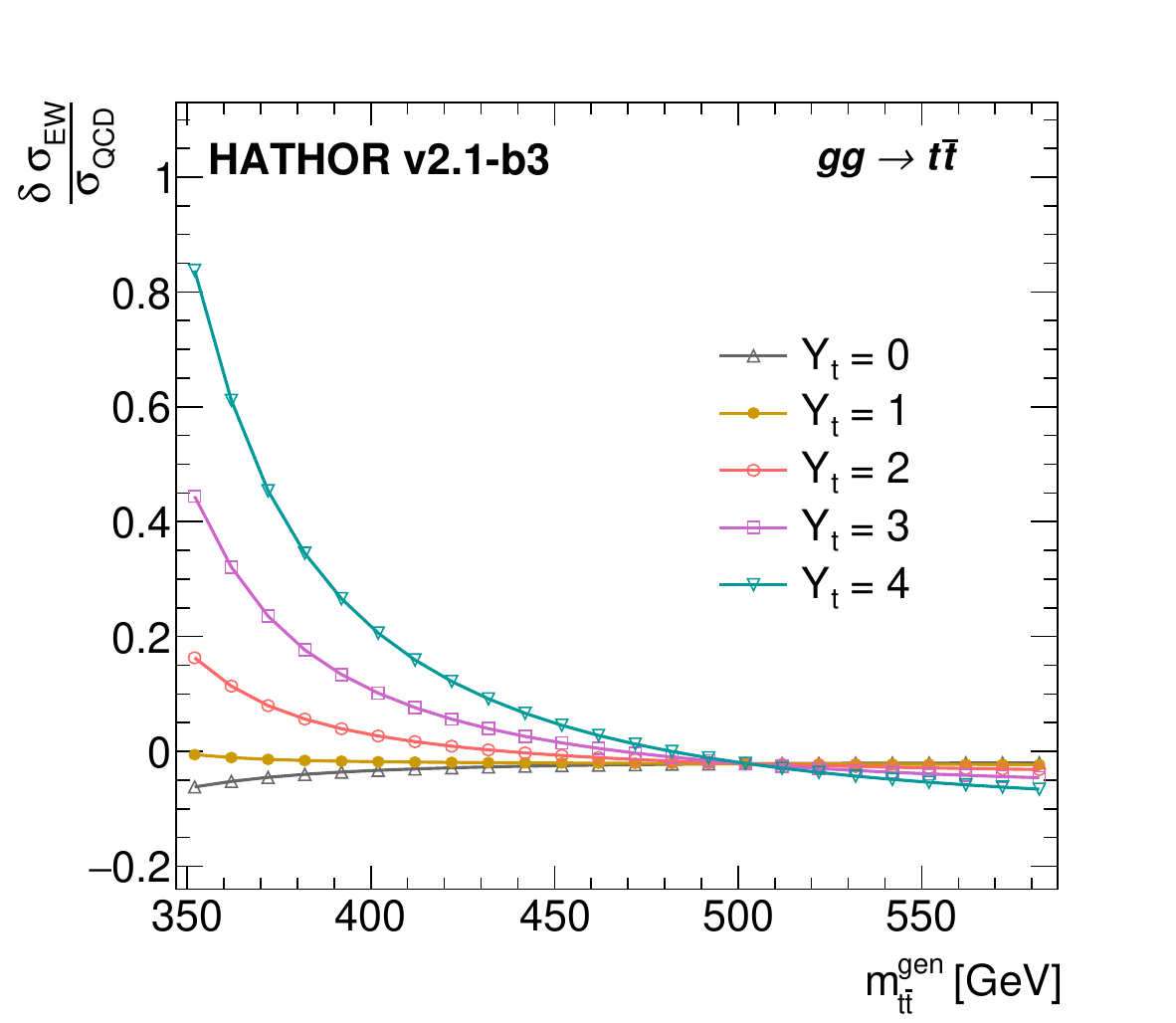}
\caption{}
\end{subfigure}
\caption{Ratio of the electroweak corrections ($\delta \sigma_\text{EW} = \sigma_\text{QCD+EW}
- \sigma_\text{QCD}$) over the leading-order QCD cross-section ($\sigma_\text{QCD}$) at parton level, as a function of $\mtt^\textrm{gen}$, the \ttbar mass calculated directly from the top quark and top antiquark four-momenta at generator level, for different \yt values from (a)
quark--antiquark annihilation and (b) gluon--gluon fusion processes,
calculated with \hathor 2.1-b3~\cite{Hathor-21-b3}. }
\label{fig:ewcorr}
\end{figure}

%
%
%
%
%
%
%

% End of text imported from the .//analysis_main/ElectroweakCorr.tex input file

%
\section{Event reconstruction and selection}
\label{sec:eventReco}

% The next lines are included from the .//analysis_main/EventReco.tex input file
\subsection{Object definitions}
\label{subsec:objdefs}

Events are required to have at least one reconstructed \pp interaction vertex with a minimum of two associated tracks with transverse momenta $\pT > 0.5~\GeV$.
The primary vertex is defined as the vertex with the highest sum of squared transverse momenta of associated tracks~\cite{ATL-PHYS-PUB-2015-026}.

Electron candidates are reconstructed from energy deposits in the electromagnetic calorimeter matched to a track in the ID~\cite{EGAM-2018-01}.
The track is required to be matched to the primary vertex, where the track longitudinal impact parameter $z_0$, must satisfy $\left|z_0 \sin{\theta}\right| < 0.5$~mm and the transverse impact parameter $d_0$ and its uncertainty $\sigma(d_0)$ must satisfy $\left| d_0/\sigma(d_0) \right| < 5$.
Electron candidates are required to be within $\left|\eta\right| < 2.47$, excluding the transition region between the barrel and the end-cap calorimeters, $1.37 < \left| \eta \right| < 1.52$.
They must satisfy $\pT > 25~\GeV$ and the \emph{Tight} likelihood identification criteria and \emph{Tight} isolation criteria~\cite{EGAM-2018-01}.

Muon candidates are reconstructed by associating tracks in the ID with tracks or track segments in the MS~\cite{MUON-2018-03}, refined through a global fit that uses hits from both sub-detectors~\cite{PERF-2015-10}.
The track longitudinal and transverse impact parameters must satisfy the requirements $\left|z_0 \sin{\theta}\right| < 0.5$~mm and $\left| d_0/\sigma(d_0) \right| < 3$, respectively.
Muon candidates must satisfy $\pT > 25~\GeV$, $\left|\eta\right| < 2.5$ and the \emph{Medium} identification and \emph{Tight} isolation criteria~\cite{MUON-2018-03}.

Jets are reconstructed from particle-flow objects~\cite{PERF-2015-09,JETM-2018-05} using the anti-\kt algorithm~\cite{Cacciari:2008gp,Fastjet} with a jet radius parameter $R = 0.4$.
The jet energy scale (JES) is calibrated using both data and simulation, as detailed in Reference~\cite{JETM-2018-05}. Additionally, the jet energy resolution (JER) in simulation is corrected to match the resolution observed in data~\cite{JETM-2018-05}.
Jets are required to satisfy $\pT > 25~\GeV$ and $\left|\eta\right| < 2.5$.
To suppress jets originating from pile-up interactions, the \emph{jet vertex tagger} (JVT) multivariate likelihood~\cite{PERF-2014-03} discriminant, using track-based variables, is applied to jets satisfying $\pT < 60~\GeV$ and $\left|\eta\right| < 2.4$, ensuring that the selected jets are matched to the primary vertex.
Jets originating from $B$-hadrons ($b$-jets) are identified using the DL1r deep neural network algorithm~\cite{FTAG-2019-07}, using a working point corresponding to a 77\% efficiency to correctly tag a $b$-jet in \ttbar{} events, providing rejection factors of approximately 6 and 134 for charm and light-flavoured jets, respectively.
Correction factors are applied to the simulated events to compensate for differences between data and simulation in the $b$-tagging efficiency for $b$-, $c$- and light-flavoured jets.

The missing transverse energy, \met{}, is defined as the magnitude of the negative vectorial sum of the transverse momenta of all calibrated electrons, muons and jets in the event.
Tracks that are not associated with any of the physics objects but originate from the primary vertex are taken into account as a soft term~\cite{JETM-2020-03}.

To avoid the double-counting of the same energy deposits in multiple objects due to reconstruction ambiguities, an overlap removal procedure is applied.
First, jets within $\DeltaR < 0.2$ of an electron are removed.
Electrons within $\DeltaR < 0.4$ of a jet are removed.
Jets with less than three ID tracks within $\DeltaR < 0.2$ of a muon are removed to avoid rejecting high-energy muons affected by calorimeter energy loss.
Finally, muons within $\DeltaR < 0.4$ of remaining jets are removed.

\subsection{Event selection}
\label{subsec:evtsel}

A number of event selection criteria are imposed to select events with a topology expected for \ttbar{} production with a single prompt lepton in the final state.

The candidate events are required to have fired one of the single-electron or single-muon triggers~\cite{TRIG-2018-05,TRIG-2018-01} and to have at least one trigger-matched electron or muon with $\pT > 27~\GeV$, ensuring that the employed triggers are fully efficient.
For the 2015 data-taking period, a leading-lepton cut of $\pT > 25~\GeV$ is applied due to lower trigger thresholds.
Events with additional electrons or muons with $\pT > 25~\GeV$ are rejected.
At least four jets with $\pT > 25~\GeV$ are required, out of which at least two of the jets are required to be $b$-tagged.

The candidate events are further categorised based on the lepton flavour
into two signal regions (SRs): $e$+jets SR and $\mu$+jets SR.
To suppress background events containing a non-prompt or a misidentified lepton, the following selection criteria are applied.
Events in the $e$+jets SR are
required to satisfy $\met > 30~\GeV$ and the reconstructed
\Wboson boson transverse
mass\footnote{$\mwt = \sqrt{(\met + E_{\text{T},\ell})^2 -
(\vec{p}_{\text{T}}^{\text{miss}} + \vec{p}_{\text{T},\ell})^2}$}
$\mwt > 30~\GeV$.  Events in the $\mu$+jets SR are required to
satisfy $\met + \mwt > 60~\GeV$.
The split of events into separate $e$+jets and $\mu$+jets SRs is performed to accurately capture the impact and the correlations of systematic uncertainties impacting leptons, jets and signal modelling. Due to the different event selections, these sources of uncertainty can impact the two SRs differently.

Additionally, for the non-prompt and misidentified leptons background estimate, two control regions (CRs), named fake-$e$ CR and fake-$\mu$ CR, are defined by inverting the \met and \mwt{} requirements of the $e$+jets SR and $\mu$+jets SR, respectively.

Finally, to suppress contribution from poorly reconstructed \ttbar{} events in SRs, additional selection criteria on event kinematics are imposed in the \ttbar{} reconstruction, described in the next section.

\subsection{\ttbar reconstruction}
\label{subsec:ttbar_reco}
As mentioned in Section~\ref{sec:intro}, the sensitivity to \yt is enhanced in the \mtt{} distribution close to the \ttbar{} production threshold. Hence, in order to measure the \ttbar invariant mass, the four-momenta of the top and the anti-top quarks are reconstructed from the event information using a dedicated algorithm which aims to find the correct assignment of reconstructed objects to the decay products of the two top quarks. If the event contains more than two $b$-jets, the two leading-\pT{} $b$-jets are considered in the reconstruction.
The hadronically decaying top quark is reconstructed first. If the event contains only two jets in addition to the two $b$-jets, they are used to reconstruct the hadronic \Wboson boson candidate. If there are three or more additional jets, the three with the highest \pT{} are selected, and the pair with an invariant mass closest to the \Wboson boson mass is used for the hadronic \Wboson boson reconstruction.
The hadronic  \Wboson{} candidate is combined with the $b$-jet yielding the mass closest to the top-quark mass to reconstruct the hadronically decaying top-quark candidate.
Thereafter, the leptonically decaying top quark is reconstructed from the kinematics of the lepton, the leftover $b$-jet and the missing transverse energy, which represents the only available information about the neutrino kinematics. To obtain the $z$-component of the neutrino momentum, a constraint on the \Wboson boson mass is exploited. Assuming energy-momentum conservation, a quadratic equation is obtained as a function of the neutrino \pz. If two real solutions are obtained, the one yielding an invariant mass of the leptonically decaying top quark closest to the top-quark mass is considered. If no real solutions are obtained, the magnitude and $\phi$ of \met{} are altered in small steps until a solution is obtained.

\begin{figure}[htbp]
\centering
\begin{subfigure}[l]{.4\textwidth}
\includegraphics[width=1.1\textwidth]{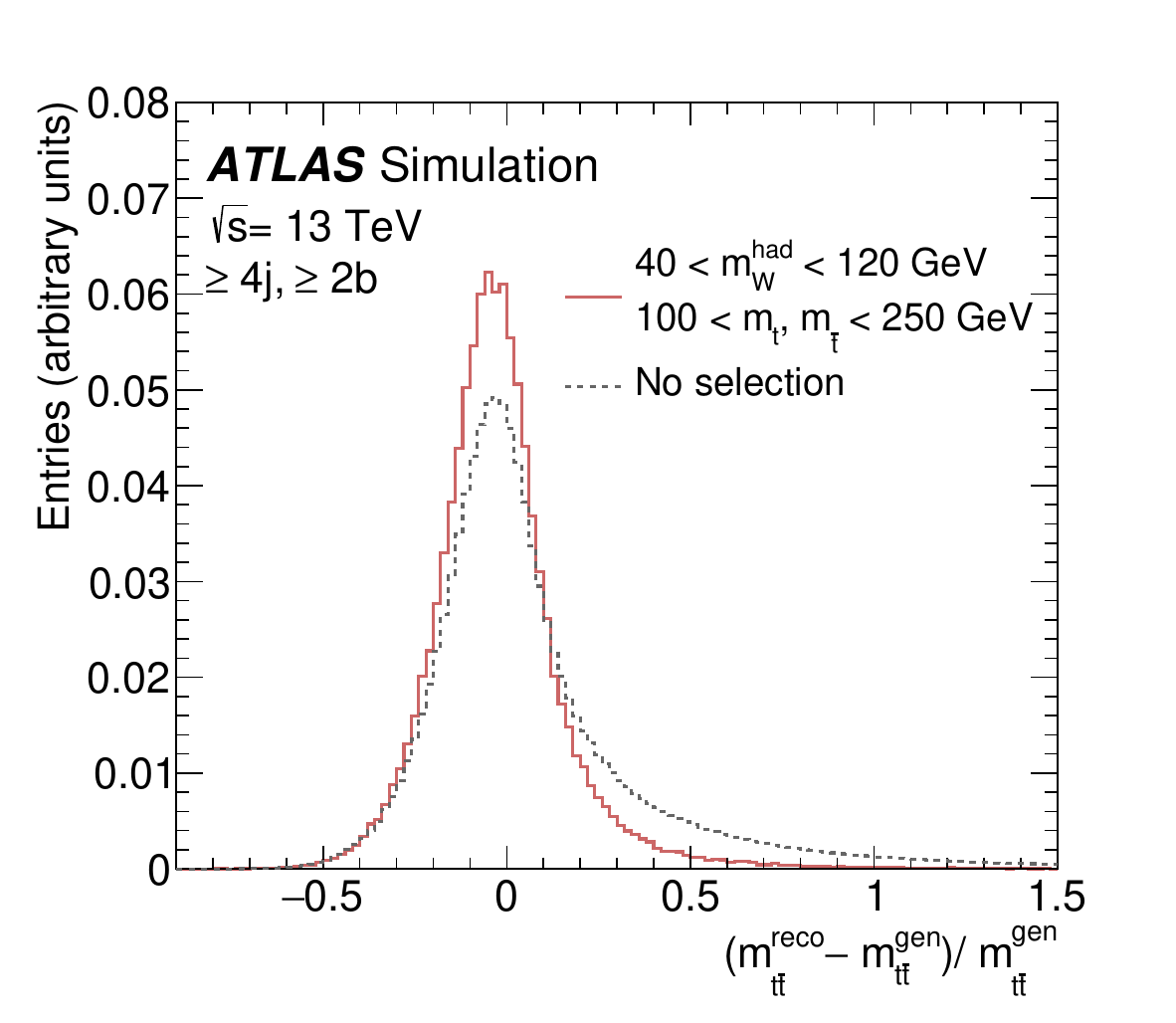}
\caption{}
\label{fig:mttbar_res}
\end{subfigure}
\hspace{4mm}
\begin{subfigure}[r]{.55\textwidth}
\includegraphics[width=1\textwidth]{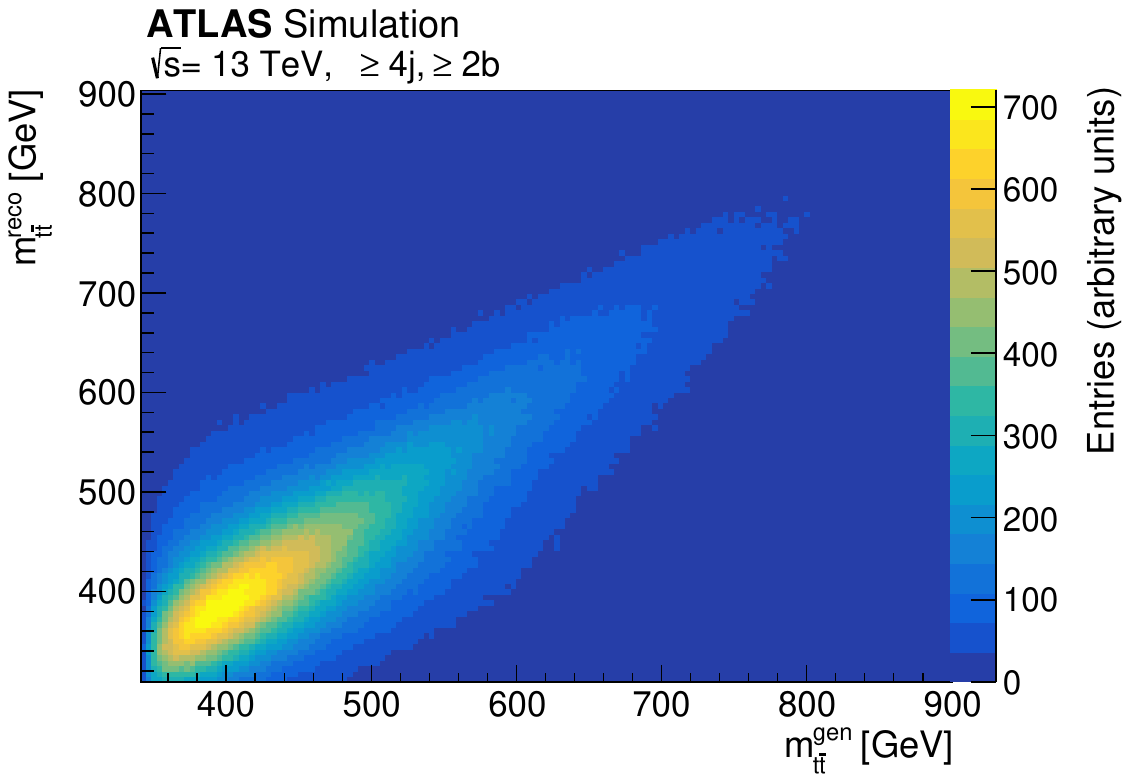}
\caption{}
\label{fig:mttbar_corr}
\end{subfigure}
\caption{Reconstructed \mtt resolution and the correlation of the reconstructed \mtt with the generator level \mtt distribution. (a) The \mtt resolution distributions after the $m_t$ and $m_{W}^\textrm{had}$ requirements, compared with the case without these selections. (b) The reconstructed \mtt after applying the selection of $m_{W}^\textrm{had} \in [40,120]~$\GeV and $m_t \in [100,250]~$\GeV, plotted against the generated \mtt.}
\label{fig:mttbar_reco}
\end{figure}

The resolution of the reconstructed mass, defined as the ratio of the
difference between the reconstructed ($\mtt^\textrm{reco}$) and the
generated ($\mtt^\textrm{gen}$) \ttbar masses\footnote{$\mtt^\textrm{gen}$ represents the \ttbar{} mass calculated directly from the top quark and top antiquark four-momenta at generator level.}, over the generated \ttbar mass value, is shown by a dashed line in Figure~\ref{fig:mttbar_res}. The distribution is centred around zero but exhibits a positive tail, making the average reconstructed \mtt larger than the generated value.
The presence of this tail may be partially attributed to incorrect jet assignments. A detailed jet-to-parton association study is performed to mitigate the selection of such events, resulting in requirements on the hadronic
$W$ boson mass (\mWhad) and on the hadronic and leptonic top quark masses. After selecting the events with $\mWhad \in [40,120]$~\GeV and $m_t \in [100,250]$~\GeV, an improvement in the resolution can be seen, as shown by the distribution with a solid line in Figure~\ref{fig:mttbar_res}.
The selection does not bias the \mtt distribution in the threshold region and consequently does not cause a bias in the \yt extraction.
Figure~\ref{fig:mttbar_corr} shows the correlation of the resultant reconstructed \mtt with the generated \mtt for events satisfying the specified requirements on $\mWhad$ and $m_t$.

\subsection{Non-prompt and fake leptons background}
\label{sub:fakes}

% The next lines are included from the .//analysis_main/FakeBkg.tex input file
The background originating from events resembling the signal topology with
either a non-prompt or misidentified lepton (referred to as \emph{fake}
lepton) is poorly modelled in the simulation.  Therefore, the fake-lepton
contribution is estimated with the data-driven \emph{matrix
method}~\cite{EGAM-2019-01}, using events with relaxed (\emph{loose})
lepton selection criteria and utilising measurements of efficiencies of the
\emph{loose} leptons to satisfy baseline (\emph{tight}) selection criteria.
The efficiencies are binned in lepton \pt{}, where the prompt lepton
efficiencies are determined from the nominal \ttbar{} simulation in the
SRs.  The fake-lepton efficiencies are determined from data in the fake
CRs, where the contribution of prompt leptons, estimated from simulation, is
subtracted from data. For both
\emph{loose} electrons~\cite{EGAM-2018-01} and \emph{loose}
muons~\cite{MUON-2018-03}, the isolation criterion is dropped.
Additionally, \emph{loose} electrons are required to satisfy a relaxed
\emph{LooseAndBLayerLH} identification~\cite{EGAM-2018-01}, while
\emph{loose} muons have a relaxed $\left| d_0/\sigma(d_0) \right| < 7$
criterion.

Due to the $\geq 2b$-jet requirement, which leads to a very high signal purity, the fake-lepton background is very small and thus there is a substantial contribution of prompt leptons in the fake CRs. Therefore, any mis-modelling of the prompt lepton contribution can have a significant impact on the fake-lepton efficiency.
In particular, the \ttbar{} simulation is known to mis-model the lepton \pt{} distribution.
This is mitigated, only for the purpose of the fake-lepton estimate, by reweighting the \ttbar{} lepton \pt{} based on the background-subtracted data in the SRs, ignoring the fake-lepton contribution, and applying the reweighting to the lepton \pt{} distribution in the fake CRs.
A closure test is performed, where the \ttbar{} lepton \pt{} reweighting is repeated after including the final fake-lepton estimate. No significant change in the prompt lepton contribution in the fake lepton CRs is observed.

Furthermore, a free correction factor (CF) to the prompt lepton efficiency is introduced in the fake CRs to account for differences between the modelling of loose and tight prompt leptons.
The value of the CF is determined such that the estimated fake-lepton \mwt{} distribution is flat around the $W$ boson mass in the SRs, indicating that the prompt background was subtracted correctly.
The central value of the CF is close to unity and a systematic variation on the CF is considered, resulting in a variation of up to 20\% on the fake-lepton \mtt{} distribution.

The resulting fake-lepton background constitutes a very small fraction of approximately 0.7\% $\pm$ 0.3\% of the total expected signal and background contribution in the SRs.

% End of text imported from the .//analysis_main/FakeBkg.tex input file

% End of text imported from the .//analysis_main/EventReco.tex input file

%
%
%
%
%
\section{Systematic uncertainties}
\label{sec:syst}

% The next lines are included from the .//analysis_main/Systematics.tex input file
The measurement is impacted by systematic uncertainties, ranging from experimental uncertainties, signal and background modelling uncertainties and uncertainties due to the finite amount of MC events.
For systematic uncertainties defined by both ``up'' and ``down'' variations, a symmetrised uncertainty is obtained by taking the arithmetic average of the magnitudes of the two variations, unless otherwise specified.
For one-sided systematic uncertainties, the ``up'' uncertainty is mirrored to obtain the ``down'' variation.
The systematic variations are smoothed to reduce the impact of statistical fluctuations due to finite amount of generated Monte Carlo sample events.

\subsection{Experimental uncertainties}%
\label{sub:exp_syst}

A number of experimental uncertainties are considered which are related to object reconstruction, pile-up modelling and luminosity. They are applied to all the processes other than the non-prompt and fake-lepton background.

The uncertainty in the integrated luminosity is 0.83\%~\cite{DAPR-2021-01}.
Additionally, to match the pile-up distribution of simulation to data, a rescaling of the average number of interaction per bunch crossing is applied to the simulation.
An uncertainty in the rescaling is propagated by varying the corresponding pile-up reweighting factors.

For electrons and muons, the uncertainties in the trigger, identification, isolation and reconstruction efficiencies are included, which are derived from studies using $Z\rightarrow ee$~\cite{EGAM-2018-01,TRIG-2018-05} and $Z\rightarrow \mu\mu$ events~\cite{MUON-2018-03,TRIG-2018-01}, respectively.
Additionally, uncertainties in the electron energy scale and muon momentum scale and their resolutions are evaluated using dedicated measurements in data~\cite{EGAM-2018-01,MUON-2022-01}.

Systematic uncertainties in JES and JER are evaluated using a series of simulation-based techniques and in situ measurements~\cite{JETM-2018-05,ATL-PHYS-PUB-2022-021}.
These include uncertainty components related to the jet flavour composition, $\eta$-intercalibration, punch-through, single-particle response, calorimeter response to different jet flavours and pile-up effects. They are comprised of 30 uncorrelated JES components and 13 uncorrelated JER components.
Additionally, an uncertainty in the calibration of the efficiency of the JVT is evaluated by varying the associated correction factor~\cite{PERF-2014-03}.
The ``up'' and ``down'' JER variations are not symmetrised due to a significantly asymmetrical impact on the fitted \mtt{} distribution.

Uncertainties in the calibrations of the $b$-tagging algorithm in data are propagated by varying the tagging efficiency correction factors~\cite{FTAG-2018-01,FTAG-2020-08,FTAG-2019-02}.
These include nine, four and four components for the $b$, $c$ and light-flavoured jet calibrations, respectively, and two components related to the extrapolation to high-\pt{} jets.

The uncertainties related to the energy scale and resolution of leptons and jets are propagated to \met{}.
Additional uncertainties in \met{} arise from momentum scale and resolution uncertainties of the track-based soft term~\cite{JETM-2020-03}.

\subsection{Signal modelling uncertainties}%
\label{sub:signal_modelling_syst}

A number of signal modelling uncertainties are considered.
A normalisation uncertainty of $^{+3.6}_{-4.5}$\% on the \ttbar{} process is assigned based on the theory uncertainty of the NNLO+NNLL QCD prediction described in Section~\ref{subsec:signal}.
Uncertainties related to the QCD scale variations in the ME are estimated by varying the \muR and \muF scales in the ME calculation independently by factors 2 and 0.5.
For the uncertainty in the amount of initial-state radiation (ISR), the first component is estimated by changing $\alpha_S^{\text{ISR}}$ via the \textsc{Var3c} variation of the \PYTHIA A14 tune.
The second component of the ISR uncertainty is estimated by varying the \hdamp{} parameter from $1.5\times \mtop$ to $3\times \mtop$.
The uncertainty related to the amount of final-state radiation (FSR) is estimated by varying the \muR scale of $\alpha_S^{\text{FSR}}$ in the parton shower by factors 2 and 0.5, respectively.
The uncertainty associated with the matching of the matrix element to the parton shower is estimated by varying the \pthard{} parameter in \PYTHIA~\cite{ATL-PHYS-PUB-2023-029}.
The uncertainties in the parton shower and hadronisation model are estimated by comparing the sample generated with \POWHER~7.2.1~\cite{Bahr:2008pv,Bahr:2008pv} with the nominal \ttbar{} sample. The corresponding variation is further decomposed into a variation for events with exactly four jets and a variation for events with five or more jets, to reduce strong constraints observed otherwise.
A set of 30 eigenvariations in the PDF4LHC15 prescription~\cite{Butterworth:2015oua} are considered to estimate the uncertainties in the choice of the PDF set.
The \mtt{} threshold region is potentially sensitive to uncertainties related to the top-quark mass and decay modelling.
The uncertainty in the top-quark mass is propagated by varying its value in the MC simulation by $\pm 0.5$~\GeV.
The uncertainty in the modelling of the top-quark decay and off-shell effects is estimated by comparing the nominal \POWPY \ttbar{} sample with a \ttbar{} sample reweighted at parton-level to the \POWMADPY sample in top-quark and top-antiquark mass observables.
The uncertainty due to missing higher-order corrections in the ME is estimated by comparing a sample reweighted by a dedicated NNLO QCD prediction, described in Section~\ref{subsec:signal}, with the nominal \ttbar{} sample.

Finally, the application of the EW corrections to the \ttbar{} sample has
an associated ambiguity, whether the correction is applied in a
multiplicative or additive approach on top of the NLO QCD corrections.
Since the QCD corrections are dominantly collinear while the
electroweak corrections occur at a much higher scale, the multiplicative
approach is considered more accurate.  This is particularly true for
\ttbar production where, due to the large top-quark mass, QCD radiation
mostly originates from the initial state, while electroweak corrections
almost entirely affect the final state~\cite{Kuhn:2013zoa}. Therefore, the multiplicative approach is used consistently
and the additive approach is evaluated in a simplified way.
In the additive approach, the electroweak corrections should only be applied to the Born-level QCD
diagrams. This is technically not possible in \POWHEG, due to the additional radiation in the ME
and because of the interplay between the ME and PS matching.
Instead, an approximation is made, rescaling the EW corrections by the ratio of the NLO QCD to the Born cross-section, taken from~\cite{threshold_Sumino_2010},
which corresponds to a factor 0.75 almost independent of \mtt{}.
The difference between the EW corrections in the multiplicative and the additive approach depends on the \yt value{}.
A complete treatment of this uncertainty would require the introduction of a nuisance parameter that is dependent on the observable of interest.
Instead, the uncertainty  is defined assuming the SM prediction of $\yt = 1$.
This assumption is deemed reasonable and the resulting uncertainty conservative, given that for this analysis the multiplicative approach is better motivated than the additive approach. For $\yt=1$, this uncertainty is smaller than $1\%$, and the impact on the final result is sub-dominant.

\subsection{Background modelling uncertainties}%
\label{sub:bckg_modelling_syst}

The event selection in the analysis ensures a very low background contamination, with an expected signal purity of $93.5$\%. Therefore, background modelling uncertainties play a sub-dominant role.

The single-top quark background is the largest background in the analysis, with the $tW$-channel contributing approximately 2\% to the total expected signal and background yield.
The normalisation uncertainties considered correspond to the approximate NNLO inclusive cross-section uncertainties of 1.9\% for the $t$-channel~\cite{Campbell:2020fhf} and 3.8\% for the $s$-channel~\cite{Kidonakis:2010tc} production. A 3.6\% normalisation uncertainty is assigned to the $tW$ process, corresponding to the approximate N$^3$LO inclusive cross-section~\cite{Kidonakis:2021vob}.
The uncertainties in the amount of ISR and FSR in the single-top quark processes are taken into account using the same variations as for the \ttbar process, but treated as uncorrelated with respect to it.
These include independent variations of the \muR and \muF scales in the ME by a factor 0.5 and 2.
Another component includes changing the \PYTHIA A14 tune settings to the \textsc{Var3c} eigentune~\cite{ATL-PHYS-PUB-2014-021}.
The uncertainty in the amount of final-state radiation is estimated by varying the \muR scale of the parton-shower emissions by factors 2 and 0.5.
Additionally, the uncertainty in the treatment of the interference and overlap between $tW$ production and \ttbar{} production is estimated by comparing the $tW$ MC sample using DR scheme with MC sample using the DS scheme instead~\cite{Frixione:2008yi,ATL-PHYS-PUB-2016-020}.

For the $W$+jets production, a normalisation uncertainty of 5\% is considered, based on the NNLO QCD cross-section predictions from \textsc{MATRIX}~\cite{ATL-PHYS-PUB-2023-026}.
An additional uncertainty is evaluated by simultaneously varying the \muR and \muF scales in the ME by a factor of 0.5 and 2, respectively~\cite{ATL-PHYS-PUB-2017-006}. The factor 0.5 and 2 variations are not symmetrised due to a significantly asymmetrical impact on the \mtt{} distribution.

The uncertainties in the toponium process modelling include a conservative 100\% normalisation uncertainty and uncertainties arising from the independent variations of factor 2 and 0.5 of \muR and \muF scales in the ME. However, since no information about the angular correlations of decay products is used to exploit the expected pseudo-scalar nature of the toponium, no significant sensitivity to this process is expected.

For the fake-lepton background, a normalisation uncertainty of 50\% is considered.
Additional uncertainties include a variation of the prompt lepton efficiency CF introduced in Section~\ref{sub:fakes} and per-bin statistical uncertainties.

Finally, a normalisation uncertainty of 50\% is considered for each process of the additional sub-dominant backgrounds including $Z$+jets, dibosons and $\ttbar{}X$ ($X=Z,W,H$).

% End of text imported from the .//analysis_main/Systematics.tex input file

%
\section{Fit strategy}
\label{sec:fit}

% The next lines are included from the .//analysis_main/FitStrategy.tex input file
%
%
%
%
%

Weights derived from \hathor are used to produce \mtt{} templates for various values of \yt{}.
A profile likelihood template fit using the \textsc{TRExFitter} tool is performed, taking these templates into account using a morphing method \cite{Cranmer:2012sba,aly_2025_15642614}.
Systematic uncertainties are included in the likelihood as nuisance parameters (NPs) with Gaussian constraints.
The statistical uncertainty in the signal and background predictions is accounted for by adding a NP for each bin, assuming a Poisson constraint~\cite{Cranmer:2012sba}.

As explained in Section \ref{sec:electroweakCorr}, the dependence of the EW corrections on \yt is exactly quadratic. Hence, \ytsq is used as the parameter of interest (POI) in the fit, instead of \yt.
The resulting linear dependence of the templates on the POI simplifies the fit and avoids issues with highly non-parabolic likelihood.
In the morphing approach used, a normalisation parameter is added to the nominal ($\ytsq = 1$) template for each bin, which is parametrised by a linear function of \ytsq{}.
The per-bin linear function parametrisations are determined from the \mtt{} histograms for multiple points in \ytsq{} space, obtaining a continuous linear interpolation in \ytsq{}.

The analysis is performed separately in the electron and muon channels, using their respective histogram templates for different \ytsq values.
A split into regions with exactly 4 jets and $\geq 5$ jets was investigated; however, due to strong constraints of signal modelling uncertainties with anti-correlations across the two regions, this approach was rejected.
Figure~\ref{fig:mttbar_reco_templates} shows the corresponding \ttbar signal templates used in the fit.
In addition to the templates corresponding to $\ytsq = 0, 1, 4, 9$, histograms corresponding to negative \ytsq values are also considered.
These non-physical negative \ytsq templates
are obtained by extrapolating the linear dependence of the electroweak corrections on \ytsq{} to negative values.
These templates are introduced in the fit in order to produce a stable fit minimisation with a continuous likelihood function for the situation where the $-1\sigma$ uncertainty in the fitted \ytsq result goes further below zero.
\begin{figure}[htbp]
\centering
\begin{subfigure}[l]{.49\textwidth}
\includegraphics[width=1\textwidth]{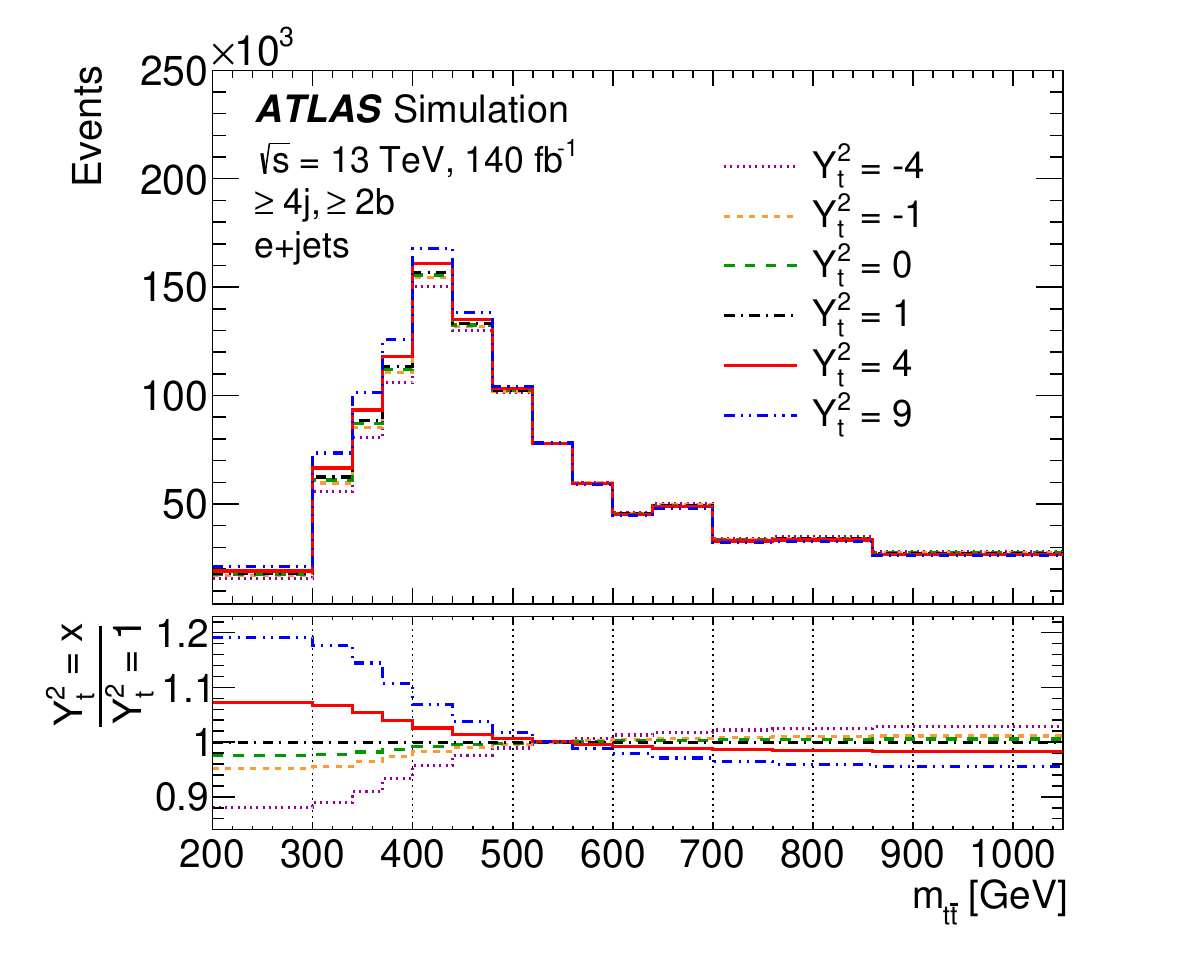}
\caption{}
\label{fig:mttbar_temp_el}
\end{subfigure}
\begin{subfigure}[r]{.49\textwidth}
\includegraphics[width=1\textwidth]{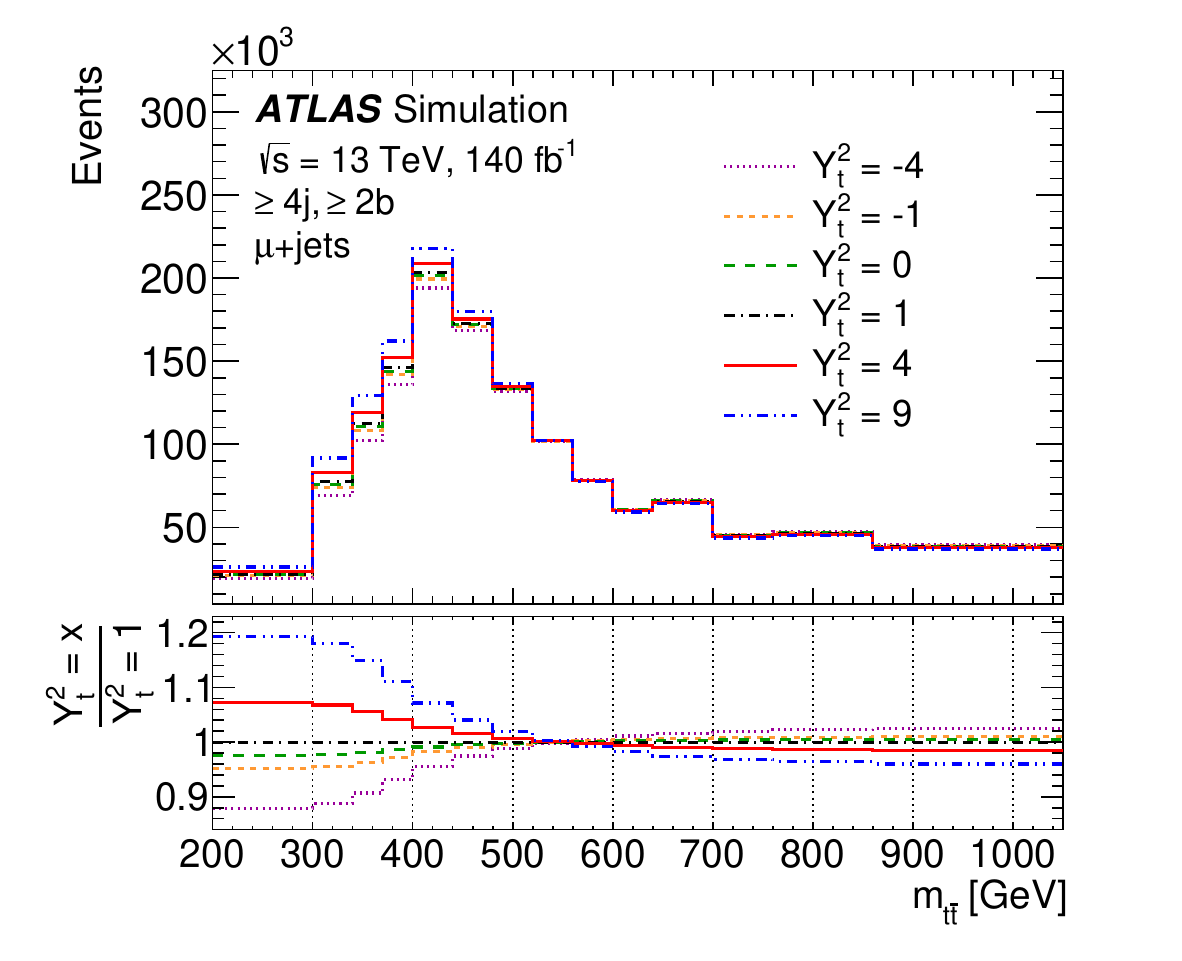}
\caption{}
\label{fig:mttbar_temp_mu}
\end{subfigure}
\caption{Reconstructed \mtt distributions corresponding to the various \ytsq templates used in the fit for (a) the electron and (b) the muon channel. The lower panel displays the ratio of the yields for each \ytsq template relative to the $\ytsq=1$ template. The negative \ytsq templates are unphysical and are used only to improve the fit stability.}
\label{fig:mttbar_reco_templates}
\end{figure}

To carefully consider the \yt-sensitive region, the fit employs an optimised binning strategy.
The fitted \mtt{} distribution consists of 14 bins in total per channel.
Two bins cover the area below the \ttbar production threshold. The binning is made finer for \mtt near the \ttbar threshold to preserve \ytsq sensitivity. %
As no steep shape dependence on \ytsq  is observed in the tail of the \mtt{} distribution, and to avoid artificial nuisance parameter constraints in the fit due to a large number of bins, away from the threshold, a wider binning scheme is adopted, with the \mtt restricted to be below 1050~\GeV.
Cross-checks are performed to test the fit dependence on the \mtt{} range by reducing the maximum \mtt{} to 700~\GeV and to 500~\GeV.
No dependence on the extracted central value of \ytsq{} is observed. The sensitivity to \ytsq is slightly reduced with the decreased \mtt{} range.
Extending the fit range up to 1050~\GeV leads to a reduction of post-fit correlations of \ytsq{} with signal modelling uncertainties, by providing additional information about their shape and acceptance effects.

%
%

%
%
%
%
%
%
%
%
%
%
%
%
%
%
%
%
%
%
%
%
%
%
%
%
%
%
%
%
%
%
%
%
%
%
%
%
%

% End of text imported from the .//analysis_main/FitStrategy.tex input file

%
\section{Results}
\label{sec:result}

% The next lines are included from the .//analysis_main/Results.tex input file
%
A comparison of the reconstructed \mtt distribution for the data to the total prediction before and after the combined fit is shown in Figure~\ref{fig:mttfinal} for the $e$+jets and $\mu$+jets SRs.
The value of \ytsq is extracted by performing a profile-likelihood fit to the data as explained in Section~\ref{sec:fit}.
The expected and observed negative log-likelihood distributions are shown in Figure~\ref{Fig:LHscan}.
The \ytsq values at $-\Delta \ln L = 0.5$ obtained from the likelihood scans give the associated $\pm1\sigma$ uncertainties.

\begin{figure}[htbp]
\centering
\begin{subfigure}[l]{.48\textwidth}
\includegraphics[width=1\textwidth]{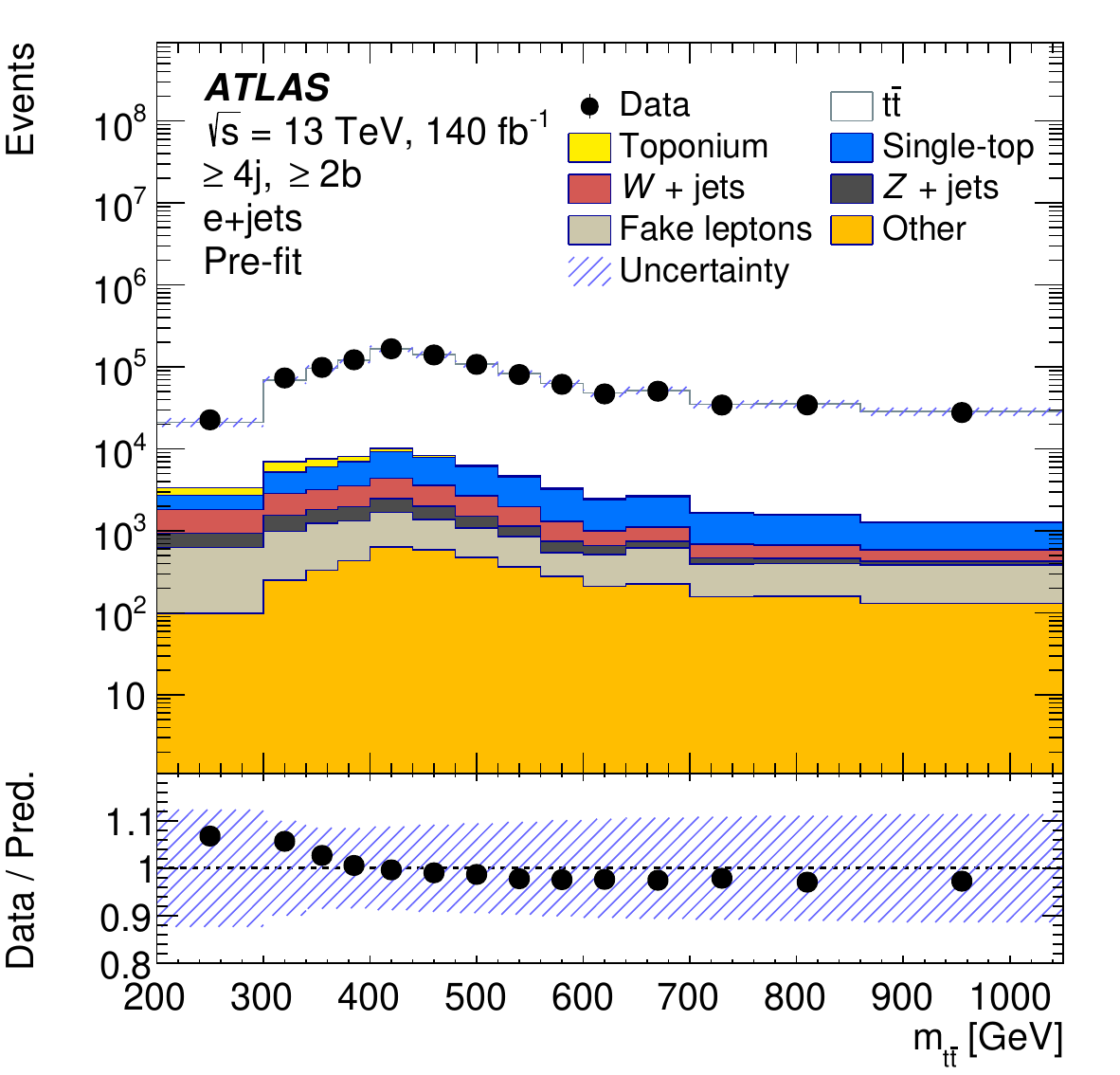}
\caption{}
\label{fig:mtt_el_pre}
\end{subfigure}
\begin{subfigure}[r]{.48\textwidth}
\includegraphics[width=1\textwidth]{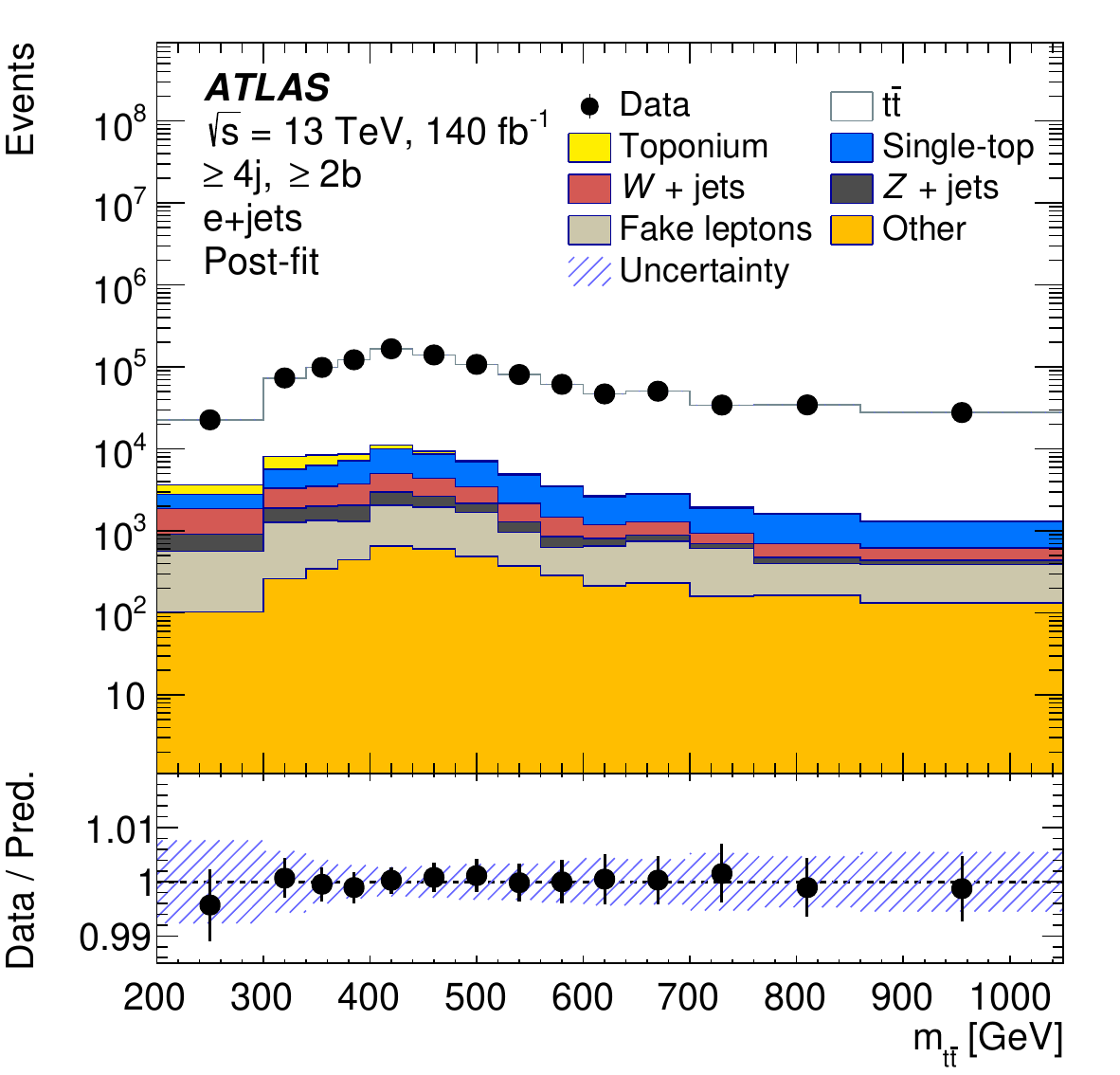}
\caption{}
\label{fig:mtt_el_post}
\end{subfigure}
\centering
\begin{subfigure}[l]{.48\textwidth}
\includegraphics[width=1\textwidth]{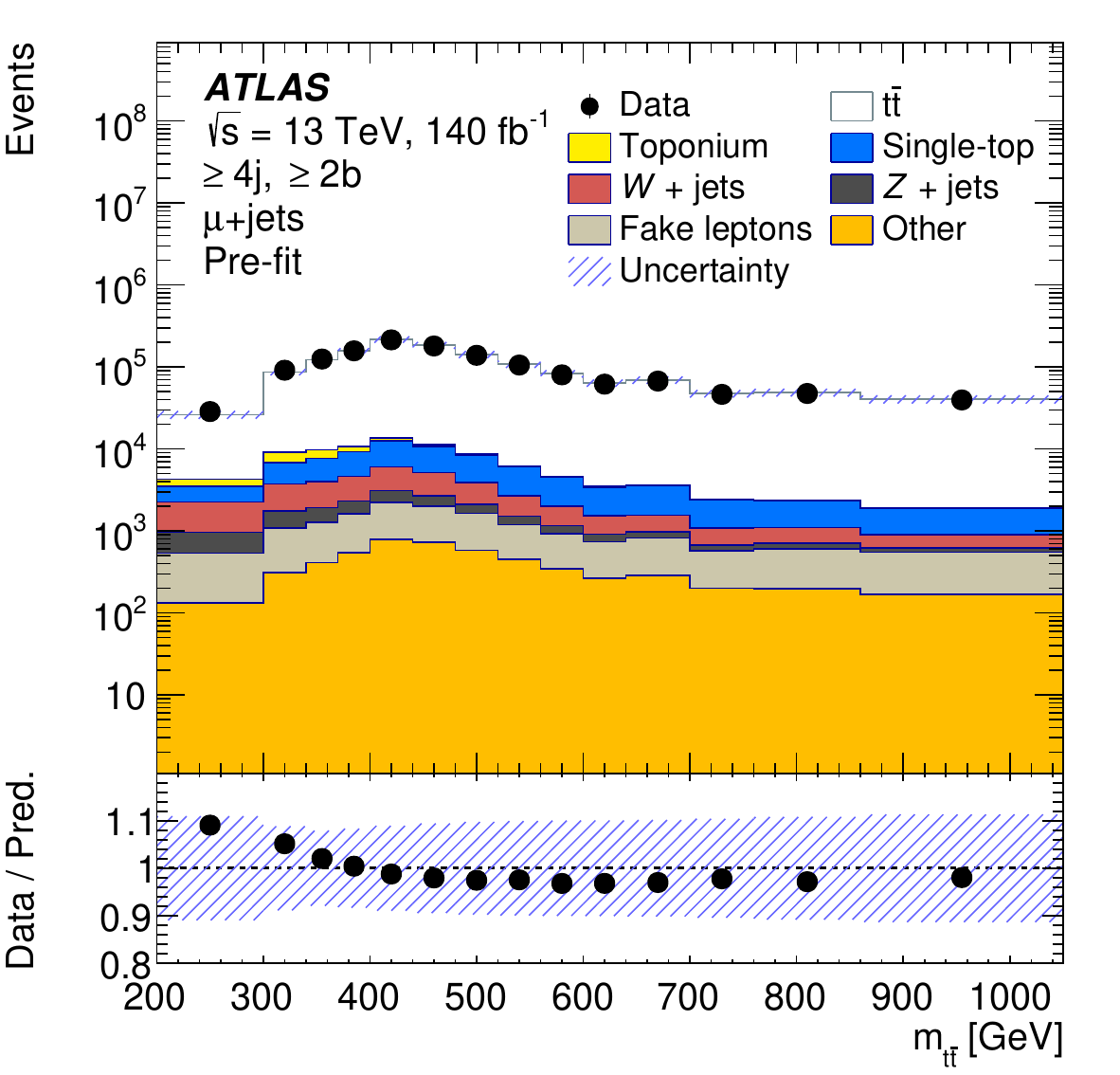}
\caption{}
\label{fig:mtt_mu_pre}
\end{subfigure}
\begin{subfigure}[r]{.48\textwidth}
\includegraphics[width=1\textwidth]{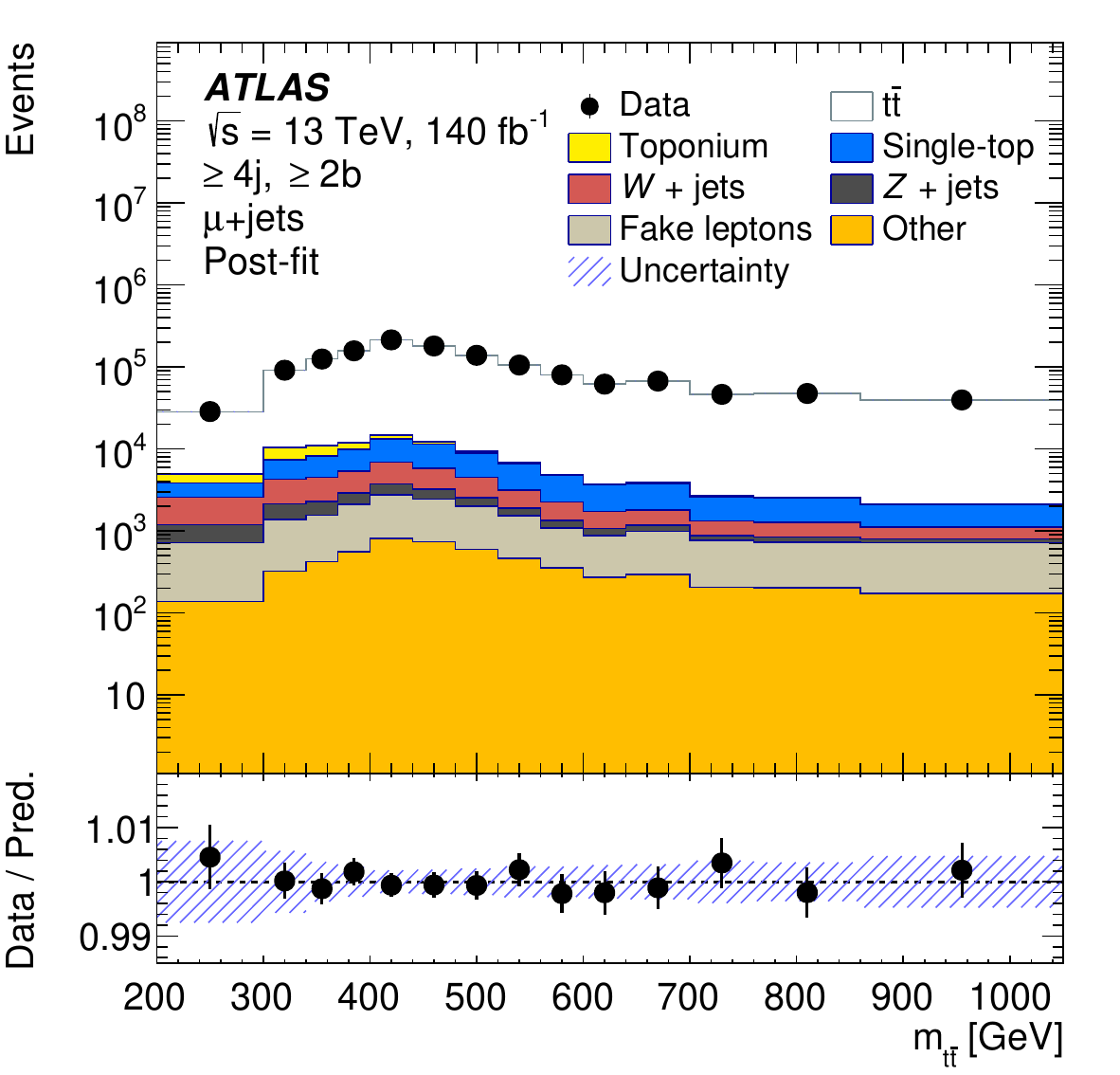}
\caption{}
\label{fig:mtt_mu_post}
\end{subfigure}
\caption{The \mtt distributions (a), (c) before the fit and (b), (d) after the combined fit, for $e$+jets and $\mu$+jets SRs, respectively. The \ttbar signal in the plots corresponds to $\yt=1$. The shaded bands represent the contribution of statistical and systematic uncertainties. The bottom panel in the figures show the ratio of the data over the total prediction.}
\label{fig:mttfinal}
\end{figure}

A summary of the \ytsq obtained from a fit to $e$+jets SR only, $\mu$+jets SR only and the simultaneous fit to both SRs (combined fit) is shown in Table~\ref{tab:result}.
The observed value of \ytsq{} is $1.3 \pm 1.7$, which is consistent with the Standard Model expectation of $\ytsq = 1$. %
A 95\% confidence level (CL) upper limit on \ytsq{} using a modified frequentist $\mathrm{CL_{s}}$ procedure~\cite{Read:2002hq} is also obtained, considering only the physical region of $\ytsq > 0$.
Subsequently, the upper limits on \yt{} for the individual SR fits as well as the combined fit are obtained, which are also summarised in Table~\ref{tab:result}.
This results in observed (expected) limits of $ \yt <  2.4 \ (2.3)$ for electrons, $ \yt <  2.2 \ (2.1)$ for muons, and $ \yt <  2.1 \ (2.1)$ for the combined result, at 95\% CL.

\begin{figure}[htbp]
\centering
\includegraphics[width=0.54\textwidth]{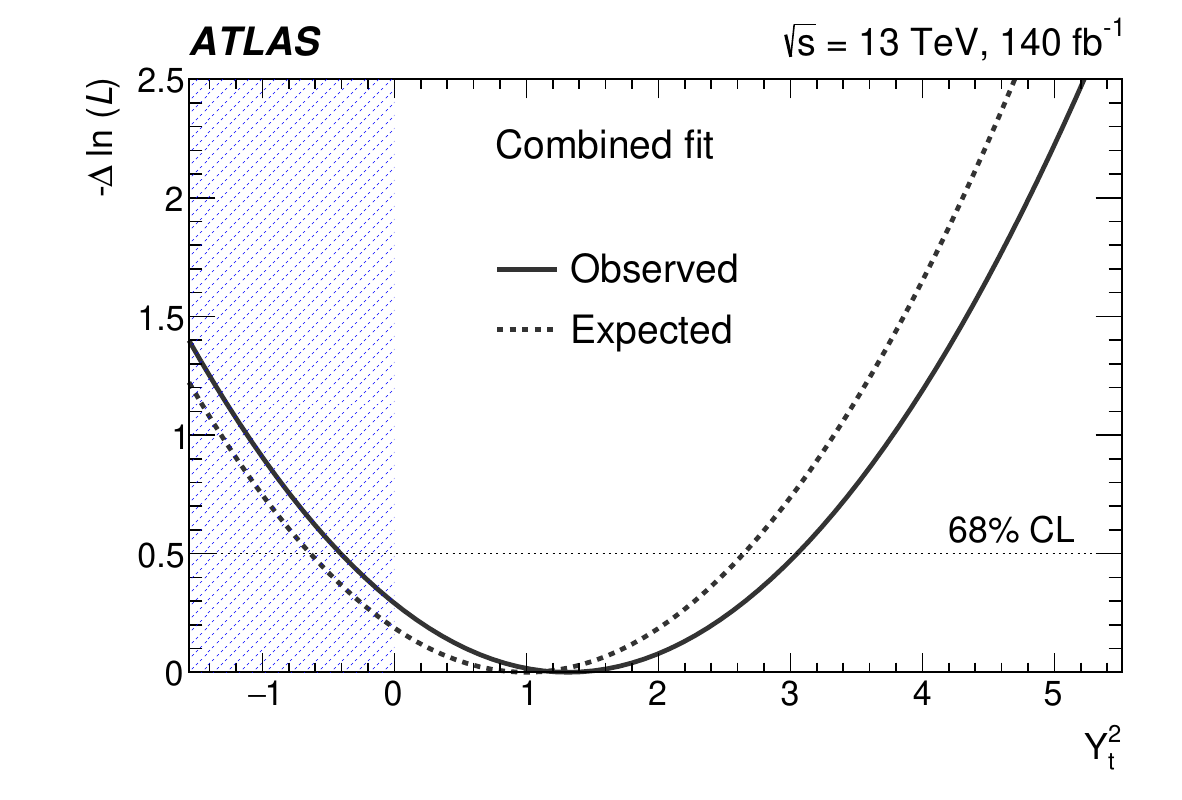}
\caption{The observed and expected negative log-likelihood \ytsq scans for the combined fit. The scan minimum shows the best-fit \ytsq. The horizontal line corresponds to the $\pm1\sigma$ uncertainty in \ytsq. The shaded region corresponds to the negative \ytsq values, which are unphysical and are only used to improve the fit stability.}
\label{Fig:LHscan}
\end{figure}

\begin{table}[ht]
\begin{center}
\caption{Summary of the results for the individual SR and the combined fit. The expected and observed best fit values for \ytsq with uncertainty and their corresponding 95\% CL upper limits on \yt are shown.}
\label{tab:result}
\begin{tabular}{c@{\hskip 0.2in}c@{\hskip 0.2in}c@{\hskip 0.2in}cc}%
\toprule
\multirow{2}{*}{\textbf{Region}} & \multirow{2}{*}{\textbf{Expected \ytsq}} & \multirow{2}{*}{\textbf{Observed \ytsq}} &  \multicolumn{2}{c}{\textbf{95\% $\mathrm{CL_{s}}$ upper limit on \yt}}\\
& & &  \textbf{\hspace{2mm}Expected} & \textbf{Observed} \\[1pt]
\toprule
$e$+jets & $1.0^{+2.1}_{-2.0}$ & $1.3^{+2.2}_{-2.1}$ & $<2.3$  & $<2.4$\\[3pt]
\toprule[0.1pt]
$\mu$+jets & $1.0^{+1.8}_{-1.8}$ & $1.1^{+1.9}_{-1.9}$ & $<2.1$ & $<2.2$ \\[3pt]
\toprule[0.1pt]
Combined & $1.0^{+1.6}_{-1.6}$ & $1.3^{+1.7}_{-1.7}$ & $<2.1$ & $<2.1$ \\[3pt]
\toprule
\end{tabular}
\end{center}
\end{table}

Figure~\ref{Fig:ranking} shows the shifts relative to pre-fit values, the constraints and the impact on \ytsq{} of the nuisance parameters (NPs) with the largest contribution to the total uncertainty in the combined fit.
The impacts of individual NPs on the measurement are extracted from the covariance matrix~\cite{pinto2024uncertaintycomponentsprofilelikelihood}.
The measurement is dominated by systematic uncertainties, where the largest impact comes from the \ttbar modelling, specifically from the \muF scale variation, followed by the JES modelling. The top quark mass and the \hdamp variation show an impact of a similar order. The strongest constraints are observed from the NPs associated with \ttbar modelling. The NNLO reweighting uncertainty is the most constrained, followed by the parton shower and hadronisation uncertainty for $\geq 5$ jets. Table~\ref{tab:groupedImpact} summarises the impact from different categories of uncertainty sources. The NPs associated with \ttbar modelling, JES, and background modelling constitute the dominant contributions to the total systematic uncertainty. Among the background modelling uncertainties, the largest contribution arises from the uncertainty in the fake-lepton background estimate.
The impact of the toponium normalisation uncertainty is negligible, at the level of $\pm 0.1$ on \ytsq{}.

\begin{figure}[htbp]
\centering
\includegraphics[width=0.65\textwidth]{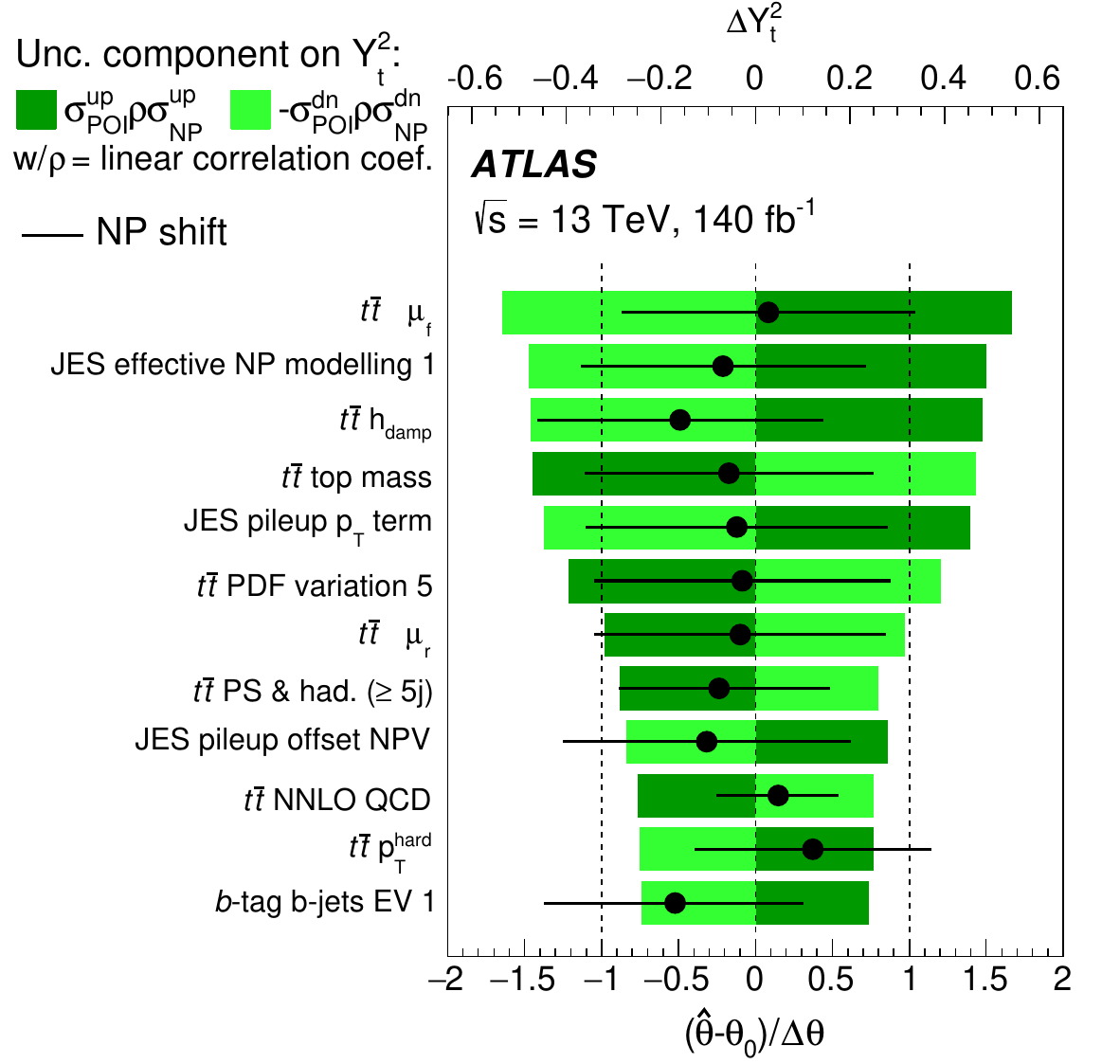}
\caption{Ranking of the most impactful NPs in the combined fit with \ytsq as the parameter of interest. The rectangular bars show the impact of each NP, given by the $\mathrm{\Delta\ytsq}$ corresponding to the top scale, estimated from the off-diagonal elements of the fit covariance matrix~\cite{pinto2024uncertaintycomponentsprofilelikelihood}.
The points show the ratio of the difference between the fitted value of the NPs $\hat{\theta}$, relative to their nominal value $\theta_0$, over the pre-fit uncertainty $\Delta{\theta}$.
The solid lines show the NP post-fit uncertainties corresponding to the bottom scale.}
\label{Fig:ranking}
\end{figure}

\begin{table}[h!]
\begin{center}
\caption{Impact of the different categories of systematic uncertainties. The impacts of the NPs on \ytsq{} in a given category are summed in quadrature. The category ``Others'' includes uncertainties in luminosity, pileup modelling, jet vertex tagger efficiency, \met,  leptons and the MC statistical uncertainty. The statistical uncertainty in \ytsq{} is obtained by subtracting the total systematic uncertainty from the total uncertainty in quadrature.
}
\label{tab:groupedImpact}
\begin{tabular}{lc}
\toprule
\textbf{Uncertainty category}& \textbf{Impact on \ytsq} \\
\toprule[0.1pt] \vspace{1mm}
\ttbar modelling & $\pm1.2$ \\ \vspace{1mm}
Jet energy scale & $\pm0.8$ \\ \vspace{1mm}
Background modelling & $\pm0.6$ \\ \vspace{1mm}
Jet energy resolution & $\pm0.4$ \\ \vspace{1mm}
$b$-tagging scale factor & $\pm0.3$ \\ \vspace{1mm}
Others & $\pm0.3$ \\
\toprule[0.1pt]
Statistical uncertainty & $\pm0.4$\\ \vspace{1mm}
Total systematic uncertainty & $\pm1.7$\\
\toprule
\end{tabular}
\end{center}
\end{table}

As discussed in Section~\ref{sub:signal_modelling_syst}, this paper adopts a multiplicative approach to incorporate electroweak corrections, while the additive approach is considered as a systematic uncertainty.
This uncertainty depends upon the value of \yt, and has been evaluated for $\yt = 1$ in the fit.
A test performed assuming this uncertainty at $\yt = 2$ results in a 6\% increase in the upper limit on \yt.

This result represents the first ATLAS measurement of \yt from the \ttbar
production threshold region, consistent with the prediction of the Standard
Model and with corresponding measurements by
CMS~\cite{CMS-TOP-17-004,CMS-TOP-19-008}. It is also consistent with all
other direct and indirect measurements of \yt.
%
%
%
%
%
%
%

%
%
%
%
%
%
%
%
%
%
%

%
%
%
%
%
%

%
%
%
%
%
%
%
%
%
%
%
%
%
%
%
%
%
%
%
%
%
%
%
%
%

%

% End of text imported from the .//analysis_main/Results.tex input file

%
%
%
\FloatBarrier
\section{Conclusion}
\label{sec:conclusion}

% The next lines are included from the .//analysis_main/Conclusion.tex input file
The top-quark Yukawa coupling strength \yt is extracted
from the differential \ttbar cross-section in events with a single lepton in the final state. This paper presents the first ATLAS measurement of \yt in the \ttbar production threshold region.
The analysis uses a $\rts = 13$~\TeV \pp collision dataset collected at the LHC and corresponding to an integrated luminosity of 140~\ifb.
The parameter of interest, \ytsq, is extracted from
a template-based profile likelihood fit to the invariant mass of the \ttbar pair,
where the \ttbar production threshold is particularly sensitive to virtual electroweak corrections,
including the Higgs boson exchange between two top quarks.
The electroweak corrections, including the free \yt parameter, are applied using \hathor %
on top of NLO QCD predictions using \POWHEGBOX-v2. %
The fitted  $\ytsq = 1.3 \pm 1.7$ is in good
agreement with the Standard Model prediction of $\ytsq = 1$ as well
as with the CMS measurements in the single-lepton~\cite{CMS-TOP-17-004} and dilepton channels~\cite{CMS-TOP-19-008}.
Normalising over the physical region, $\ytsq > 0$, yields a 95\% CL limit of $\yt < 2.1$.
The presented measurement provides a complementary approach compared to other indirect measurements with a similar sensitivity to constrain the \yt coupling strength.

% End of text imported from the .//analysis_main/Conclusion.tex input file

%

%
%
%
%
%
\section*{Acknowledgements}
%

%

% The next lines are included from the .//acknowledgements/Acknowledgements.tex input file
%
%

%
%

We thank CERN for the very successful operation of the LHC and its injectors, as well as the support staff at
CERN and at our institutions worldwide without whom ATLAS could not be operated efficiently.

The crucial computing support from all WLCG partners is acknowledged gratefully, in particular from CERN, the ATLAS Tier-1 facilities at TRIUMF/SFU (Canada), NDGF (Denmark, Norway, Sweden), CC-IN2P3 (France), KIT/GridKA (Germany), INFN-CNAF (Italy), NL-T1 (Netherlands), PIC (Spain), RAL (UK) and BNL (USA), the Tier-2 facilities worldwide and large non-WLCG resource providers. Major contributors of computing resources are listed in Ref.~\cite{ATL-SOFT-PUB-2025-001}.

We gratefully acknowledge the support of ANPCyT, Argentina; YerPhI, Armenia; ARC, Australia; BMWFW and FWF, Austria; ANAS, Azerbaijan; CNPq and FAPESP, Brazil; NSERC, NRC and CFI, Canada; CERN; ANID, Chile; CAS, MOST and NSFC, China; Minciencias, Colombia; MEYS CR, Czech Republic; DNRF and DNSRC, Denmark; IN2P3-CNRS and CEA-DRF/IRFU, France; SRNSFG, Georgia; BMFTR, HGF and MPG, Germany; GSRI, Greece; RGC and Hong Kong SAR, China; ICHEP and Academy of Sciences and Humanities, Israel; INFN, Italy; MEXT and JSPS, Japan; CNRST, Morocco; NWO, Netherlands; RCN, Norway; MNiSW, Poland; FCT, Portugal; MNE/IFA, Romania; MSTDI, Serbia; MSSR, Slovakia; ARIS and MVZI, Slovenia; DSI/NRF, South Africa; MICIU/AEI, Spain; SRC and Wallenberg Foundation, Sweden; SERI, SNSF and Cantons of Bern and Geneva, Switzerland; NSTC, Taipei; TENMAK, T\"urkiye; STFC/UKRI, United Kingdom; DOE and NSF, United States of America.

Individual groups and members have received support from BCKDF, CANARIE, CRC and DRAC, Canada; CERN-CZ, FORTE and PRIMUS, Czech Republic; COST, ERC, ERDF, Horizon 2020, ICSC-NextGenerationEU and Marie Sk{\l}odowska-Curie Actions, European Union; Investissements d'Avenir Labex, Investissements d'Avenir Idex and ANR, France; DFG and AvH Foundation, Germany; Herakleitos, Thales and Aristeia programmes co-financed by EU-ESF and the Greek NSRF, Greece; BSF-NSF and MINERVA, Israel; NCN and NAWA, Poland; La Caixa Banking Foundation, CERCA Programme Generalitat de Catalunya and PROMETEO and GenT Programmes Generalitat Valenciana, Spain; G\"{o}ran Gustafssons Stiftelse, Sweden; The Royal Society and Leverhulme Trust, United Kingdom.

In addition, individual members wish to acknowledge support from CERN: European Organization for Nuclear Research (CERN DOCT); Chile: Agencia Nacional de Investigaci\'on y Desarrollo (FONDECYT 1230812, FONDECYT 1240864, Fondecyt 3240661, Fondecyt Regular 1240721); China: Chinese Ministry of Science and Technology (MOST-2023YFA1605700, MOST-2023YFA1609300), National Natural Science Foundation of China (NSFC - 12175119, NSFC 12275265); Czech Republic: Czech Science Foundation (GACR - 24-11373S), Ministry of Education Youth and Sports (ERC-CZ-LL2327, FORTE CZ.02.01.01/00/22\_008/0004632), PRIMUS Research Programme (PRIMUS/21/SCI/017); EU: H2020 European Research Council (ERC - 101002463); European Union: European Research Council (BARD No. 101116429, ERC - 948254, ERC 101089007), European Regional Development Fund (SMASH COFUND 101081355, SLO ERDF), European Union, Future Artificial Intelligence Research (FAIR-NextGenerationEU PE00000013), Italian Center for High Performance Computing, Big Data and Quantum Computing (ICSC, NextGenerationEU); France: Agence Nationale de la Recherche (ANR-21-CE31-0022, ANR-22-EDIR-0002, ANR-24-CE31-0504-01); Germany: Deutsche Forschungsgemeinschaft (DFG - 469666862, DFG - CR 312/5-2); China: Research Grants Council (GRF); Italy: Istituto Nazionale di Fisica Nucleare (ICSC, NextGenerationEU), Ministero dell'Università e della Ricerca (NextGenEU 153D23001490006 M4C2.1.1, NextGenEU I53D23000820006 M4C2.1.1, NextGenEU I53D23001490006 M4C2.1.1, SOE2024\_0000023); Japan: Japan Society for the Promotion of Science (JSPS KAKENHI JP22H01227, JSPS KAKENHI JP22H04944, JSPS KAKENHI JP22KK0227, JSPS KAKENHI JP24K23939, JSPS KAKENHI JP24KK0251, JSPS KAKENHI JP25H00650, JSPS KAKENHI JP25H01291, JSPS KAKENHI JP25K01023); Norway: Research Council of Norway (RCN-314472); Poland: Ministry of Science and Higher Education (IDUB AGH, POB8, D4 no 9722), Polish National Science Centre (NCN 2021/42/E/ST2/00350, NCN OPUS 2023/51/B/ST2/02507, NCN UMO-2019/34/E/ST2/00393, UMO-2022/47/O/ST2/00148, UMO-2023/49/B/ST2/04085, UMO-2023/51/B/ST2/00920, UMO-2024/53/N/ST2/00869); Portugal: Foundation for Science and Technology (FCT); Spain: Ministry of Science and Innovation (MCIN \& NextGenEU PCI2022-135018-2, MICIN \& FEDER PID2021-125273NB, RYC2019-028510-I, RYC2020-030254-I, RYC2021-031273-I, RYC2022-038164-I); Sweden: Carl Trygger Foundation (Carl Trygger Foundation CTS 22:2312), Swedish Research Council (Swedish Research Council 2023-04654, VR 2021-03651, VR 2022-03845, VR 2022-04683, VR 2023-03403, VR 2024-05451), Knut and Alice Wallenberg Foundation (KAW 2018.0458, KAW 2022.0358, KAW 2023.0366); Switzerland: Swiss National Science Foundation (SNSF - PCEFP2\_194658); United Kingdom: Royal Society (NIF-R1-231091); United States of America: U.S. Department of Energy (ECA DE-AC02-76SF00515), Neubauer Family Foundation.

%
%

% End of text imported from the .//acknowledgements/Acknowledgements.tex input file

%
\clearpage
\printbibliography
\clearpage
\input{atlas_authlist}
%

%
%

%

%
%
%
%

\end{document}

%% file: atlas_authlist.tex
% ATLAS Collaboration author list
% Reference date of TOPQ-2021-11 is 2025-08-16
% Author list last updated on date 03-NOV-25
% Data extracted on 03-Nov-2025 for paper reference TOPQ-2021-11
% at 3:14pm
 
\begin{flushleft}
\hypersetup{urlcolor=black}
{\Large The ATLAS Collaboration}

\bigskip

\AtlasOrcid[0000-0002-6665-4934]{G.~Aad}$^\textrm{\scriptsize 103}$,
\AtlasOrcid[0000-0001-7616-1554]{E.~Aakvaag}$^\textrm{\scriptsize 17}$,
\AtlasOrcid[0000-0002-5888-2734]{B.~Abbott}$^\textrm{\scriptsize 122}$,
\AtlasOrcid[0000-0002-0287-5869]{S.~Abdelhameed}$^\textrm{\scriptsize 118a}$,
\AtlasOrcid[0000-0002-1002-1652]{K.~Abeling}$^\textrm{\scriptsize 55}$,
\AtlasOrcid[0000-0001-5763-2760]{N.J.~Abicht}$^\textrm{\scriptsize 49}$,
\AtlasOrcid[0000-0002-8496-9294]{S.H.~Abidi}$^\textrm{\scriptsize 30}$,
\AtlasOrcid[0009-0003-6578-220X]{M.~Aboelela}$^\textrm{\scriptsize 45}$,
\AtlasOrcid[0000-0002-9987-2292]{A.~Aboulhorma}$^\textrm{\scriptsize 36e}$,
\AtlasOrcid[0000-0001-5329-6640]{H.~Abramowicz}$^\textrm{\scriptsize 156}$
% ...
\bigskip
\\

% ...
$^{17}$Department for Physics and Technology, University of Bergen, Bergen; Norway.\\
% ...
$^{30}$Physics Department, Brookhaven National Laboratory, Upton NY; United States of America.\\
% ...
$^{36}$$^{(a)}$Facult\'e des Sciences Ain Chock, Universit\'e Hassan II de Casablanca;$^{(b)}$Facult\'{e} des Sciences, Universit\'{e} Ibn-Tofail, K\'{e}nitra;$^{(c)}$Facult\'e des Sciences Semlalia, Universit\'e Cadi Ayyad, LPHEA-Marrakech;$^{(d)}$LPMR, Facult\'e des Sciences, Universit\'e Mohamed Premier, Oujda;$^{(e)}$Facult\'e des sciences, Universit\'e Mohammed V, Rabat;$^{(f)}$Institute of Applied Physics, Mohammed VI Polytechnic University, Ben Guerir; Morocco.\\
% ...
$^{45}$Physics Department, Southern Methodist University, Dallas TX; United States of America.\\
% ...
$^{49}$Fakult\"{a}t Physik, Technische Universit{\"a}t Dortmund, Dortmund; Germany.\\
% ...
$^{55}$II. Physikalisches Institut, Georg-August-Universit\"{a}t G\"ottingen, G\"ottingen; Germany.\\
% ...
$^{103}$CPPM, Aix-Marseille Universit\'e, CNRS/IN2P3, Marseille; France.\\
% ...
$^{118}$$^{(a)}$New York University Abu Dhabi, Abu Dhabi;$^{(b)}$United Arab Emirates University, Al Ain; United Arab Emirates.\\
% ...
$^{122}$Homer L. Dodge Department of Physics and Astronomy, University of Oklahoma, Norman OK; United States of America.\\
% ...
$^{156}$Raymond and Beverly Sackler School of Physics and Astronomy, Tel Aviv University, Tel Aviv; Israel.\\
% ...
$^{a}$ Also at Affiliated with an institute formerly covered by a cooperation agreement with CERN.\\
$^{b}$ Also at An-Najah National University, Nablus; Palestine.\\
$^{c}$ Also at Borough of Manhattan Community College, City University of New York, New York NY; United States of America.\\
$^{d}$ Also at Center for Interdisciplinary Research and Innovation (CIRI-AUTH), Thessaloniki; Greece.\\
$^{e}$ Also at Centre of Physics of the Universities of Minho and Porto (CF-UM-UP); Portugal.\\
$^{f}$ Also at CERN, Geneva; Switzerland.\\
% ...

\end{flushleft}

% Created with Glance <Atlas.Glance@cern.ch>